\documentclass[aps,prl,twocolumn,showpacs,superscriptaddress,10pt]{revtex4-1}

\usepackage{amsmath}
\usepackage{amsfonts}
\usepackage{amssymb}
\usepackage{graphicx}
\usepackage{color}
\usepackage{xcolor}
\usepackage{mathtools}
\usepackage{mathrsfs}
\usepackage[normalem]{ulem}

\newcommand{\beq}{\begin{equation}}
\newcommand{\eeq}{\end{equation}}

\begin{document}
%
%\title{Reshaping dynamical critical points \\
%with out-of-equilibrium quantum fluctuations}
%
\title{Chaotic dynamical ferromagnetic phase \\ induced by non-equilibrium quantum fluctuations}
\author{Alessio Lerose}
\affiliation{SISSA --- International School for Advanced Studies, via Bonomea 265, I-34136 Trieste, Italy}
\affiliation{INFN --- Istituto Nazionale di Fisica Nucleare, Sezione di Trieste, I-34136 Trieste, Italy}
\author{Jamir Marino}
\affiliation{Institut f\"{u}r Theoretische Physik, Universit\"{a}t zu K\"{o}ln, D-50937 Cologne, Germany}
\author{Bojan \v{Z}unkovi\v{c}}
\affiliation{Department of Physics, Faculty of Mathematics and Physics, University of Ljubljana, Jadranska 19, 1000 Ljubljana, Slovenia}
\author{Andrea Gambassi}
\affiliation{SISSA --- International School for Advanced Studies, via Bonomea 265, I-34136 Trieste, Italy}
\affiliation{INFN --- Istituto Nazionale di Fisica Nucleare, Sezione di Trieste, I-34136 Trieste, Italy}
\author{Alessandro Silva}
\affiliation{SISSA --- International School for Advanced Studies, via Bonomea 265, I-34136 Trieste, Italy}

\begin{abstract}

We investigate the robustness of a dynamical phase transition against quantum fluctuations by studying the 
impact of a ferromagnetic nearest-neighbour spin interaction in one spatial
dimension on the non-equilibrium dynamical phase diagram of the fully-connected
quantum Ising model. 
In particular, we focus on the transient dynamics {after a quantum quench} 
and study the pre-thermal state via a combination of analytic time-dependent spin-wave theory and numerical methods based on matrix product states.
We find that, upon increasing the strength of the quantum fluctuations, the dynamical critical point fans out into a chaotic dynamical phase within which  the asymptotic ordering is characterised by 
strong sensitivity to the parameters and initial conditions. 
We argue that such a phenomenon is general, as it arises from the  impact of quantum fluctuations on the mean-field out of equilibrium  dynamics of any system which exhibits a broken discrete symmetry.
\end{abstract}


%05.30.Rt	    Quantum phase transitions
%05.70.Ln	Nonequilibrium and irreversible thermodynamics
%64.60.Ht	Dynamic critical phenomena
% 75.10.Jm	Quantized spin models, including quantum spin frustration
% 05.30.Jp	Boson systems
% 71.10.Fd	Lattice fermion models (Hubbard model, etc.)
% 64.60.ae   RG theory
% 64.70.qj    dynamics and criticality

\date{\today}
\maketitle

%\agcomm{From the discussion below it is unclear what happens in the thermodynamic limit; perhaps we should mention this.}

\emph{Introduction} ---
Throwing dice on the table floor is a prototypical random (or pseudorandom) process~\cite{Dicebook}. Its aleatory nature is a consequence of a few ingredients: the die, initially out of equilibrium, dissipates its energy rolling on the table, and hence relaxes onto one of  few possible equilibrium configurations. In this work we  show that  the same ingredients play an important role in the physics of quantum many-body systems undergoing a Dynamical Quantum Phase Transition (DQPT), leading to the emergence of an intriguing chaotic dynamical phase once the collective dynamics of these systems gets damped by quantum fluctuations.
%Recent progress in  controlling and manipulating the dynamics of ultra-cold  %gases~\cite{Greiner2002a,*Greiner2002b,*Bloch2008,*Trotzky2008,*Cheneau2012,*Kauf16,*Jorg15,*rauer17} 
%and trapped ions \cite{hess17} 
%
%
%has revived the interest in quantum statistical physics out of equilibrium. In particular,
%it is currently possible to realise in the lab models of condensed-matter physics, with an exquisite 
%control of their parameters and geometries, making possible to use time-resolved measurements to access the non-equilibrium evolution of relevant observables,
%posing new challenges to our
%understanding of {these systems} \cite{PolkovnikovRMP, *Lamacraft2012,*Eisert2015a}.

DQPTs are among the most interesting  phenomena occurring in quantum many-body systems after a sudden change of the system parameters (\emph{quantum quench}) \cite{PolkovnikovRMP,*Lamacraft2012,*Eisert2015a}, a type of process which can be realized both with ultra-cold  gases~\cite{Greiner2002a,*Greiner2002b,*Bloch2008,*Trotzky2008,*Cheneau2012,*Kauf16,*Jorg15,*rauer17} 
and trapped ions \cite{hess17}.
{Such DQPTs~\cite{Sciolla2011,Smacchia2014,Maraga2016,Halimeh2016, *Halimeh2017,*Gambassi2011} are characterised by the vanishing of a non-equilibrium order parameter (accompanied by  critical scaling  behaviour~\cite{Sciolla2013, Chiocchetta2015}) and to be 
%depending whether the quantum model under study is quenched respectively below or above an associated dynamical critical point.%
%~In particular, in systems with local interactions, the scaling of dynamical correlation (or response) functions provide diagnostics of different phases:  quenching the system close to the %dynamical critical point, un-equal time correlation functions manifest a scaling behaviour reminiscent of aging in classical critical dynamics~\cite{Janssen1989, *Calabrese2005,  %Maraga2015}, while quenching across  dynamical critical points, imprints real-time scaling behaviour compatible with dynamical coarsening \cite{Maraga2015,*Chandran2013}. 
%
%For shallow quenches  critical exponents controlling  scaling behaviour are dictated by a quantum non-equilibrium critical point, while for deep quenches  critical exponents are in %relation with the former through the conventional quantum-to-classical dimensional correspondence~\cite{Sachdevbook}.~The non-equilibrium phase diagram of DPT  therefore   %mirrors in structure the  one of finite temperature quantum phase transitions ~\cite{Sachdevbook}, with a line of classical dynamical critical points terminating into a quantum one when %the quench amplitude crosses over from deep to shallow~\cite{Chiocchetta2017}.
%
%This class of DPT is at the core of the study of this work and should be 
distinguished  from those signalled by non-analyticities in the temporal evolution of the Loschmidt echo~\cite{Heyl2013} (see Ref.~\cite{Bojan2016b} for connections between the two notions). They
not only provide a genuine instance of classical and quantum criticality out of equilibrium
\cite{Sciolla2013, Chandran2013, *Chiocchetta2015, Chiocchetta2017},  
but demonstrate also the emergence of intermediate stages of relaxation with nontrivial time-dependent fluctuations and dynamics \cite{Smacchia2014}.}
A DQPT separates ``phases'' characterised by qualitatively different quasi-stationary states 
\cite{Sciolla2011,Gambassi2011}, anomalous coarsening \cite{Maraga2015}, aging \cite{Janssen1989, *Calabrese2005,  Maraga2015, Chiocchetta2017},  as well as by a non-trivial dynamical evolution of observables and 
their fluctuations \cite{Chandran2013, Smacchia2014,Maraga2016,Bojan2016b,Halimeh2016, *Halimeh2017}. 
Due to the lack of spatial and temporal collective scales upon approaching a DQPT, they display features reminiscent of equilibrium critical points.

%%%%%%%%%OLD INTRODUCTION%%%%%%%%%
%These dynamical phase transitions 
%
%{(DPT, to be distinguished \cite{Bojan2016b} from the notion in Ref.~\cite{Heyl2013})}
%
%not only provide a genuine instance of classical and quantum criticality out of equilibrium
%\cite{Sciolla2013, *Chandran2013, *Chiocchetta2015, Chiocchetta2017},  
%but demonstrate also the emergence of intermediate stages of relaxation with nontrivial time-dependent fluctuations and dynamics \cite{Smacchia2014}.
%
%A DPT separates ``phases'' characterised by qualitatively different quasi-stationary states 
%\cite{Gambassi2011}, anomalous coarsening \cite{Maraga2015}, aging \cite{Janssen1989, *Calabrese2005,  Maraga2015, Chiocchetta2017},  as well as by a non-trivial dynamical evolution of observables and 
%their fluctuations \cite{Smacchia2014,Maraga2016,Bojan2016b,Halimeh2016, *Halimeh2017}. 
%
%Due to the lack of spatial and temporal collective scales upon approaching a DPT, they display features reminiscent of equilibrium critical points.
%
%

\begin{figure}[t!]
\includegraphics[width=9cm]{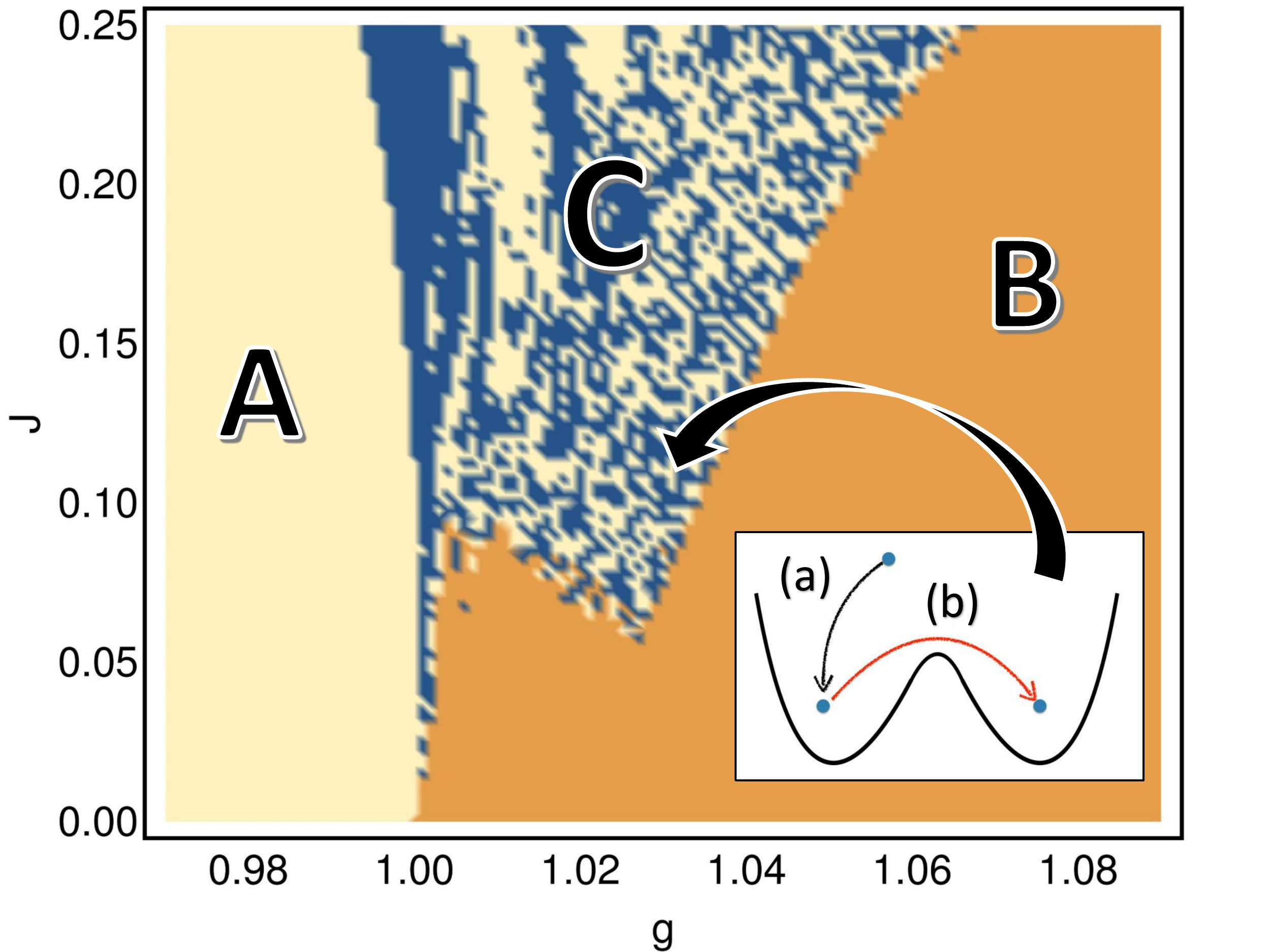}
\caption{%(Color online) 
Dynamical phase diagram of the model in Eq.~\eqref{eq:ham} after a quantum quench starting from the ferromagnetic ground state with $g=0$ and positive expectation value $S_x(0)=1 $ of the global magnetization, in the plane of the post-quench value $g$ of the transverse field and $J$ of the nearest-neighbour coupling. %For each point $(g,J)$ we numerically integrate the evolution equations of 
as obtained from the time-dependent spin-wave theory, see the main text and  \cite{SM}  (here $N=100$). We consider here the range of values of $g$ and $J$ within which the low-density  spin-wave expansion is applicable, and units are chosen such that $\bar{\lambda}\equiv \lambda+J=1$.
The color of each point of the diagram is determined by the value of long-time average $\bar S_x$ of $S_x$: light yellow for $\bar S_x>0$, orange for $\bar S_x=0$, and blue for $\bar S_x<0$. 
Regions A and B correspond to the dynamic ferromagnetic and paramagnetic phase, respectively, of the mean-field model ($J=0$). 
Upon increasing $J$ at fixed $g$ close to the mean-field critical point, i.e., $g\simeq\bar{\lambda}$,  a new \emph{chaotic dynamical ferromagnetic phase} C arises, exhibiting relaxation from an initial paramagnetic behavior to symmetry-broken sectors  (process (a) in the inset) sometimes followed by assisted hopping between the two sectors with opposite signs of $\bar S_x$ (process (b) in the inset). See Fig.~\ref{fig2} for an illustration of the dynamics in region C.% \jm{The boundaries separating the various phases in the $(g,J)$-diagram become sharper as $N$ increases.}
}
\label{fig1}
\end{figure}

DQPTs are expected to be strongly affected by quantum fluctuations: recent investigations beyond mean-field approximations \cite{Schiro2011,Sandri12, Sciolla2013,Peronaci2015} 
showed that these fluctuations influence, e.g., the early stages of the evolution \cite{Maraga2015, Chiocchetta2015, Chiocchetta2017}. 
%
%
%While these studies focused primarily on cases in which quantum fluctuations are controlled by varying the spatial dimensionality of the system, 
In this work we demonstrate a more dramatic effect of fluctuations  on the dynamics of the order parameter, which induces a qualitative modification of the dynamical phase diagram, in particular close to the dynamical critical point.  
We %focus -- for the sake of definiteness -- on spin systems described by a quantum Ising model with infinite-range interactions and an additional %nearest-neighbour one-dimensional
study the non-equilibrium dynamics of an infinite-range (mean-field) ferromagnetic system perturbed by additional
short-range interaction terms, which rule the strength of quantum fluctuations. We {show}  %demonstrate 
that the dynamical %paramagnetic and ferromagnetic 
phases are robust, whereas the impact of non-equilibrium quantum fluctuations makes the dynamical critical point open up in a 
novel \emph{chaotic dynamical phase} where the dynamics are reminiscent of that of a coin toss: The asymptotic stationary state displays a finite magnetization whose positive or negative sign is highly sensitive to initial conditions and system parameters, as we show in Fig.\ref{fig1}.
%characterised 
%by a high sensitivity of the final magnetisation to the system parameters and the initial conditions:
%arbitrarily close values of them give rise to orbits of the spin order parameter that
%
%In particular, the dynamics of the spin order parameter is regulated by a set of effective, non-linear 
%
%
%equations of motion, and trajectories starting from arbitrarily close initial points 
%depart from each other  
%diverge 
%at long times and become uncorrelated, in a fashion reminiscent of classical chaos~\cite{Gutzwillerbook}. 
%\jm{Such a phenomenon is by no means a peculiarity of the model under consideration, but actually pertains to a very wide class of systems%, the only crucial ingredients being 1) the presence of a broken discrete symmetry and 2) the presence of any interaction with a non-trivial spatial structure that breaks the mean-field limit
%~\cite{Long}.}

\emph{The model}  --- In this work, for the sake of definiteness, we focus on a fully-connected quantum Ising ferromagnet in a transverse magnetic field $g$, in the presence of additional nearest-neighbor couplings in one spatial dimension, governed by the Hamiltonian 
\begin{equation}
\label{eq:ham}
H=-\frac{\lambda}{N}\sum^N_{i,j=1}\sigma^x_i\sigma^x_j-g\sum^N_{i=1}\sigma^z_i-J\sum^N_{i=1}\sigma^x_i\sigma^x_{i+1},
\end{equation}
where $\sigma^\alpha_i$ are the standard Pauli matrices at lattice site $i$.
%
%This model is exactly solvable in the limits $\lambda=0$ and $J=0$:
%{$H$ is exactly solvable both for $\lambda=0$ and $J=0$ :}
%
%For $\lambda=0$, it reduces to the textbook quantum Ising chain, with an equilibrium quantum critical point at zero temperature and $g=J$  \cite{Sachdevbook}. 
In the limit $J\to0$, % instead, 
$H$ maps to the  exactly solvable Lipkin-Meshkov-Glick (LMG) model~\cite{LMG} and  displays both a quantum critical point in equilibrium~\cite{dutta01,*knap13}   at $g=2\lambda$ and a DQPT after a quench~\cite{Sciolla2011, BoyanMF}, with the longitudinal global magnetization $S_x(t)$ --- %where 
{with} $S_\alpha \equiv\langle \sum_i\sigma_i^\alpha \rangle/N$  ---
{being} % playing the role of
the dynamical order parameter of the DPT. 
For example, for quenches starting from the ferromagnetic ground state at $g_0=0$, 
$S_x(t)$ evolves periodically with a period set by the post-quench values $g$ and $\lambda$ of the couplings.
In particular, the time average $\bar{S}_x=\lim_{T\to\infty}\int^T_0 {\rm d}t\, S_x(t)/T$ vanishes for $g>\lambda$ 
{(because the oscillations of $S_x(t)$ are symmetric around zero)} 
while it does not for $g<\lambda$ ({because the oscillations do not change the sign of $S_x$}), 
%as a consequence of asymmetric oscillations 
%within the region $\sign (S_x)=\sign (S_x(t=0)$), 
corresponding to the dynamically paramagnetic and ferromagnetic ``phases'', respectively.
At the  dynamical critical point, $g=\lambda$, the order parameter decays exponentially to zero with 
$S_x(t)\sim {\rm e}^{-gt}$  for $t\gg g^{-1}$. 
Within mean-field theory, which is an exact treatment of the LMG model in the thermodynamic limit~\cite{BoyanMF},  the DQPT can be rationalized \cite{Gambassi2011,Sciolla2013,BoyanMF} in terms of the motion of a classical particle with position $S_x(t)$ in an effective, double-well even potential ${\cal U}(S_x)$.  
If $S_x(t=0)$ is such that ${\cal U}(S_x(0)) > {\cal U}(0)$, then $S_x(t)$ explores both wells and $\bar S_x=0$;  otherwise the motion is localized within one well and $\bar S_x\neq 0$.

In this work we study  how this mean-field non-equilibrium phase diagram is affected by quantum fluctuations. 
While for $J=0$ all spins {perform} %move according to 
a coherent collective motion, turning on a short-range perturbation is expected to damp the persistent classical oscillations of $S_x(t)$, altering the features of the mean-field evolution and inducing relaxation towards a stationary and eventual thermal state.

In order to address these questions we develop a spin-wave theory in the reference frame aligned with the instantaneous average total spin, with the spin-coherent state in this direction representing the instantaneous spin-wave vacuum. {While for $J=0$ the length of the total spin is constantly maximal, i.e., $\lvert S(t) \rvert\equiv 1$, in the presence of a small short-range perturbation $J\ne0$ a finite density $\epsilon(t)$ of spin-wave excitations is generated by the precessing collective spin, yielding $\lvert S(t) \rvert=1-\epsilon(t)$}. As long as $\epsilon(t) \ll 1$, the
non-linear, inelastic scattering among spin-waves is negligible and thermalization is expected to occur at longer times. Accordingly, the temporal regime with  $\epsilon(t)\ll 1$, within which the mean-field motion receives correction from having $J\neq 0$ while keeping its non-equilibrium features (e.g., the DQPT), can be qualified as  being pre-thermal, in analogy with similar cases~\cite{Langen2016}.

%
%\jm{In particular, we treat the non-linearities introduced by the short-range perturbation at the leading quadratic order in an Holstein-Primakoff (HP) expansion, which is controlled by %the smallness of  spin wave density $\epsilon(t)$  generated during non-equilibrium dynamics; as customary, the approximation ceases to be valid at times when $\epsilon(t)$ becomes %sizeable, which  then requires to include higher, non-quadratic contributions from the HP expansion, leading eventually to thermalization.}~Accordingly, the temporal regime with  
%$%%\epsilon(t)\ll 1$, within which the mean-field motion receives \jm{Gaussian} corrections from having $J\neq 0$ while keeping its non-equilibrium features (e.g., the DPT), can be %qualified as  being pre-thermal, in analogy with similar cases~\cite{Langen2016}.

%%%%%%%%%%%%%%%%%%%%%%%%%%%%%%%
%%%%%%%%%%%%%%%%%%%%%%%%%%%%%%%
%%%%%%%%%%%%%%%%%%%%%%%%%%%%%%%
%%%%%%%%%%%%%%%%%%%%%%%%%%%%%%%

\emph{Outline of results} --- In the presence of quantum fluctuations, one would expect the collective motion of $S_x(t)$ to be damped by the generation of spin-wave excitations with a finite rate, leading to the breakdown of the approximation $\epsilon(t)\ll 1$. (Throughout the paper we fix energy units such that $\bar{\lambda}\equiv \lambda+J=1$.) 
We find, instead, that for small $J \lesssim 0.25$,  $\epsilon(t)$ always saturates, implying that  the dynamical paramagnetic and ferromagnetic phases indicated by A and B, respectively, in Fig.~\ref{fig1} are stable.
In particular, $S_x(t)$ approximately oscillates with a period which is perturbatively close to the mean-field one. 
A numerical analysis based on the time-dependent variational principle (MPS-TDVP) \cite{LOV15,HLO14} 
indicates that this stability extends to larger values of $J$
% ,e.g., $J=0.67$,
where the spin-waves density is no longer small.
This implies that $H$ inherits the dynamical phase diagram of the classical case with $J=0$. However, the presence of spin-wave excitations makes the dynamical critical point at $\lambda =g$ for $J=0$, fan out in a 
\emph{chaotic dynamical ferromagnetic phase}  denoted by C in Fig.~\ref{fig1}.
Within C, %this phase, 
the non-equilibrium quantum fluctuations in the form of spin-waves act effectively as a self-generated bath responsible for the localization of the system, initially with $S_x>0$, into one of the 
two wells of ${\cal U}$, with either sign of $S_x$ and for the possible hopping of the collective spin $S_x$ between the two of them; these processes are sketched as (a) and (b), respectively, in the inset of Fig.~\ref{fig1}. 
%
%In the chaotic dynamical ferromagnetic phase, 
The strong sensitivity of the long-time ferromagnetic ordering to the values of $g$, $\lambda$, and $J$ and of the initial data can be regarded as a signature of a collective chaotic behavior. 
The stability of this picture upon increasing $N$ in both the analytic and the numerical approaches leads us to conclude that such behavior carries over to the thermodynamic limit.

\emph{Time-dependent spin-wave theory} --- We now briefly outline the non-equilibrium spin-wave {theory at %which constitutes 
the core of this} work \cite{SM}. 
We first introduce a time-dependent reference frame $\mathcal{R}=(\hat{X},\hat{Y},\hat{Z})$ in the spin space, with its $\hat{Z}$-axis following the collective motion of $\vec{S}(t)$. 
{The change of frame}
%This transformation 
is implemented by the time-dependent global rotation operator $V\big(\theta(t),\phi(t)\big)=\exp\left(-i\phi(t) \sum_i \sigma_i^z/2\right)\exp\left(-i\theta(t) \sum_i \sigma_i^y/2\right)$ parameterized by the angles $\theta(t)$ and $\phi(t)$, which 
%will be consistently 
are eventually determined in such a way that $S_X(t)\equiv S_Y(t)\equiv 0$.
For $J=0$, when $H$ is a function of the total spin only,
this requirement translates into a closed pair of (classical) ordinary differential equations, the solution of which 
determines the evolution of the order parameter $S_x(t)=\sin\theta(t)\cos\phi(t)$ \cite{BoyanMF}.
For %As soon as 
$J\ne0$, the additional short-range interaction renders $H$ a function of not only the total spin, i.e., the $k=0$ Fourier mode of the spins, but also of all the $k$-modes of the spins, which now contribute to the dynamics. 
In order to make the equations of motion tractable and to set up a 
{systematic expansion,} %approach, 
we introduce the canonically conjugated  spin-wave  variables $q_i$ and $p_i$ at site $i$ with respect to the instantaneous $\hat{Z}$-axis via the Holstein--Primakoff (HP) transformation
\begin{equation}
\label{eq:approxH-P}
\frac{\sigma_i^X}{2} \simeq \sqrt{s} \, q_i ,\quad \frac{\sigma_i^Y}{2}  \simeq \sqrt{s} \, p_i, \quad \frac{\sigma_i^Z}{2}  = s - \frac{q_i^2+p_i^2-1}{2}.
\end{equation}
We then express all the spin operators in $H$, see Eq.~\eqref{eq:ham}, in terms of the spin-wave coordinates in Fourier space $\tilde{q}_k$, $\tilde{p}_k$, and retain up to quadratic terms in the spatial fluctuations modes $(\tilde{q}_k,\tilde{p}_k)$ with $k\ne0$ (i.e., we neglect collisions among spin-waves).  After averaging the Heisenberg equations of motion of the spins over the nonequilibrium state~\cite{SM}, we find that 
$\theta$ and $\phi$
%the spherical angles 
evolve according to (recall $\bar{\lambda}\equiv\lambda+J$)
\begin{equation}
\label{eq:vacuummotiondelta}
\left\{
\begin{split}
\frac{d\theta}{dt} =&\, 4 \left[ \bar{\lambda}   \rho(t) - J \delta^{pp}(t) \right]  \sin\theta \cos\phi \sin\phi  \\
                         & \quad\quad \quad +4J \delta^{qp}(t) \, \cos\theta \sin\theta \cos^2\phi ,\\
\frac{d\phi}{dt}  =& -2g  + 4 \left[\bar{\lambda}  \rho(t) - J  \delta^{qq} (t)\right] \cos\theta \cos^2\phi  \\
                                          &\quad\quad\quad  +4J  \delta^{qp} (t)\, \sin\phi \cos\phi   ,
\end{split}
\right.
\end{equation}
where $\delta^{\alpha\beta}(t) \equiv  \sum_{k\ne0} \Delta^{\alpha\beta}_k \cos k/(Ns)$ with $\alpha, \beta \in \{p,q\}$ is the quantum ``feedback'' given by the %two-point 
correlation functions of the spin-waves,
\begin{equation}
\begin{split}\label{eq:Delta}
\Delta^{qq}_k (t) &\equiv  \left\langle \tilde{q}_k(t) \tilde{q}_{-k}(t) \right\rangle, \quad
\Delta^{pp}_k (t) \equiv  \left\langle \tilde{p}_k(t) \tilde{p}_{-k}(t) \right\rangle, \\
\Delta^{qp}_k (t) &\equiv \left\langle \tilde{q}_k(t) \tilde{p}_{-k}(t) + \tilde{p}_k(t) \tilde{q}_{-k}(t) \right\rangle/2.
\end{split}
\end{equation}
The relevance of these spin-wave excitations is {controlled by the quantity} %measured by the quantity
\begin{equation}
\label{eq:eps}
\epsilon(t)\equiv\frac{1}{N/2} \sum_{k\ne0} (\Delta^{qq}_k + \Delta^{pp}_k -1)/2,
\end{equation}
i.e., by %which is 
the total number of spin-waves divided by $N/2$.
In Eq.~\eqref{eq:vacuummotiondelta}, $\rho(t) = 1-\epsilon(t)$ is the ratio between the expectation value of the modulus of the total spin of the system and its maximal value $N/2$, which is conserved by the dynamics only when $J=0$~\cite{BoyanMF}. 
The evolution of $\Delta^{\alpha\beta}_k$ in Eq.~\eqref{eq:Delta} is {ruled} %governed 
by a  system of linear differential equations {involving} %depending on also on 
$\theta(t)$ and $\phi(t)$ \cite{SM}.
The quadratic approximation is justified as long as the density of {excited} spin-waves %in the system
is small, i.e., $\epsilon(t)\ll1$.
For a quench starting from the spin coherent state fully polarized in the $\hat{x}$ direction, considered here (i.e., from $g=0$) 
%
%,
%
the initial data of Eqs.~\eqref{eq:vacuummotiondelta} are 
%%
%$\theta(t=0) =  \pi/2$, $\phi(t=0) =0$ with $\Delta^{qq}_k(t=0) = \Delta^{pp}_k(t=0) = 1/2$, and $\Delta^{qp}_k(t=0) = 0$  for %all 
%$k\ne0$; in particular, $\epsilon(t=0)=0$.
{$\theta(0) =  \pi/2$, $\phi(0) =0$ with $\Delta^{qq}_k(0) = \Delta^{pp}_k(0) = 1/2$, and $\Delta^{qp}_k(0) = 0$  for 
$k\ne0$; in particular, $\epsilon(0)=0$ (note that at time $t=0$ the mobile $\hat{Z}$-axis is aligned with the fixed $\hat{x}$ direction).}
Equation~\eqref{eq:vacuummotiondelta} includes the feedback terms $\delta^{\alpha\beta}(t)$ 
from quantum fluctuations, which both ``dress'' the value of $\bar\lambda$ and generate new terms of pure quantum origin in addition to the classical mean-field
dynamics corresponding to $J=0$ in Eq.~\eqref{eq:vacuummotiondelta}.

\emph{Nonequilibrium quantum phase diagram} --- 
Via a joint numerical integration of Eq.~\eqref{eq:vacuummotiondelta} and of 
the evolution equations (see Eqs.~(26) in Ref.~\cite{SM}) of $\Delta^{\alpha\beta}_k$ in Eq.~\eqref{eq:Delta}, for a range of post-quench values of $g$ and $J$, we obtained the dynamical ``phase'' diagram portrayed in Fig.~\ref{fig1}. 
In particular, for each  integration, we compute the direction $(\theta(t), \phi(t))$ of the total spin $\vec S$ and the  
density $\epsilon(t)$ of spin-waves, verifying that the latter always settles around a small value at long times within the range of parameters considered here.
Then, we compute the long-time average $\bar{S}_x$ of $S_x(t)$ %the dynamical order parameter 
and color the corresponding point in light yellow if $\bar{S}_x>0$, in orange if $\bar{S}_x = 0$, and in blue if  $\bar{S}_x<0$.
The results of this procedure, in Fig.~\ref{fig1}, shows that the two dynamical ferromagnetic and paramagnetic phases  present for $J=0$ survive %even for
at $J>0$: for $g\ll\bar{\lambda}$ the order parameter has non-zero time-average (its value being perturbatively close to the mean-field one), while for $g\gg \bar{\lambda}$ it vanishes for all the values of $J$ within the range considered here. 
In particular, the persistent oscillations of $S_x$, characteristic of the mean-field solution, are not wiped out by the spin-wave bath, which does not produce a significant noise during the pre-thermal stage of dynamics and leaves the overall motion perturbatively close to perfect coherence for all observation times.

\begin{figure}
\includegraphics[width=8.55cm]{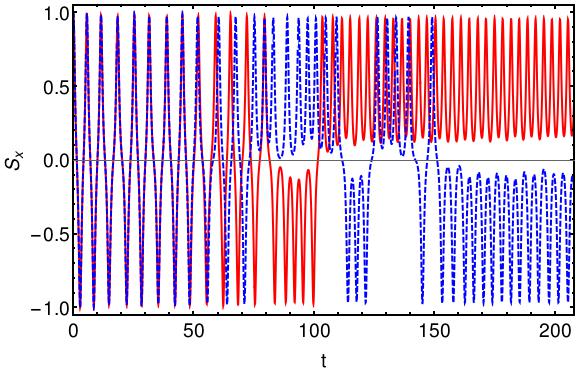}
\caption{%(Color online) 
Evolution of the order parameter $S_x(t)$ in the chaotic dynamical ferromagnetic phase 
(indicated by C in Fig.~\ref{fig1}) for $\bar{\lambda}\equiv \lambda+J=1$, $g=1.03$, with $J=0.1$ (solid red) and $J=0.1001$ (dashed blue), i.e., two very close points in the non-equilibrium phase diagram, located at the ending point of the black arrow in Fig.~\ref{fig1}, as obtained from the time-dependent spin-wave theory, see the main text and  \cite{SM} (here $N=200$). %The inset refers to the case $J=0.1$ and shows the corresponding evolution of the %corresponding 
%spin-wave density $\epsilon(t)$. 
%
The %dynamical order parameter 
magnetization $S_x(t)$  initially displays a paramagnetic behavior, 
{with %slightly decreasing amplitude in time corresponding to 
a gradual loss of energy in favor of the creation of spin-waves, witnessed by a growth of $\epsilon(t)$. % in the inset. 
This 
%triggers the trapping of the orbit 
makes the orbit fall}
into one of the two ferromagnetic wells, corresponding to process (a) of Fig.~\ref{fig1}.  %(red dashed line: the instantaneous amplitude of the oscillations of $S_x(t)$ is locally minimal, and the corresponding value $\epsilon(t)$ has a local maximum). 
However, it might later reabsorb %some 
spin-waves and hop to the opposite %ferromagnetic 
sector, corresponding to  process (b) of Fig.~\ref{fig1}. 
%(green dashed line: the order parameter hops from the region $S_x<0$ to $S_x>0$, and $\epsilon(t)$ has correspondingly a local minimum). 
%
%
The two lines are practically on top of each other during the initial  paramagnetic transient, but show %undergo 
completely different fates at the onset of the critical process (a) and they eventually end up into 
{distinct wells}.
%opposite ferromagnetic sectors. 
{(In both cases $\epsilon(t)$ grows from $\epsilon(0)=0$ to values around $0.04$ in the final stage.)} % -- see also \cite{SM}).
Such extreme sensitivity illustrates the ``mosaic'' appearance of the region C in Fig.~\ref{fig1}.}
\label{fig2}
\end{figure}

Near the dynamical transition point $g\simeq \bar{\lambda}$, the system becomes extremely sensitive to the non-equilibrium quantum fluctuations, realizing  a peculiar intermediate phase, the existence of which is intimately related to the conservation of the energy.
In a typical point of this region, 
%\ag{Within C,} 
the dynamics of $S_x$ is characterized by the processes illustrated by the inset of Fig.~\ref{fig1} and by Fig.~\ref{fig2}: the decay from a transient paramagnetic behavior to one of the two ferromagnetic sectors, 
possibly followed by one or more hops between them.  
Typically, after an initial transient with the energy of the macroscopic total spin slightly above the barrier ${\mathcal U}(0)$ separating the two ferromagnetic wells of the effective potential ${\mathcal U}$, the production of spin-waves causes the dynamics to get trapped within one of the two. %This is accompanied by an increase of the spin-wave density $\epsilon(t)$, reported in the inset of Fig.~\ref{fig2}. 
The system thus shows ferromagnetic order at long times, though it might occasionally hop to the opposite well, %sector, 
assisted by the absorption of energy from the spin-waves. The asymptotic sign of $S_x(t)$, %magnetization, 
and hence of $\bar S_x$,  sensitively depends  on the specific values of the parameters in a large part of this novel ferromagnetic region (indicated by C in Fig.~\ref{fig1}), implying a collective chaotic character of the dynamics within it, as illustrated in Fig.~\ref{fig2}.
{Unlike the quantum critical cone emanating from equilibrium quantum critical points at finite temperature~\cite{Sachdevbook}, the boundaries of region C are expected to be
sharp both towards the ferromagnetic and the paramagnetic phases, signalling two transitions expected to be characterized by diverging time scales.}
%in region C of Fig.~\ref{fig1}  the order parameter falls always asymptotically in one of the two dynamical ferromagnetic  minima and  therefore the system is not expected to display %any critical scaling, contrary to what occurs within the equilibrium quantum critical cone.}
%For further details see \cite{SM}.
%
%%
%%
%%%%%%%%%%%%%%%%%%%%%%%%%%%%%%%%%%%%%%%%%%%%
%%%%%%%%%%%%%%%%%%%%%%%%%%%%%%%%%%%%%%%%%%%%
%%%%%%%%%%%%%%%%%%%%%%%%%%%%%%%%%%%%%%%%%%%%
%%%%%%%%%%%%%%%%%%%%%%%%%%%%%%%%%%%%%%%%%%%%
%%
%%
%%
%%%%%%%%%%%%%%%%%%%%%%%%%%%%%%%%%%%%%%%%%%
%%%%%%%%%%%%%%%%%%%%%%%%%%%%%%%%%%%%%%%%%%
\begin{figure}[t!]
\includegraphics[width=8.5cm]{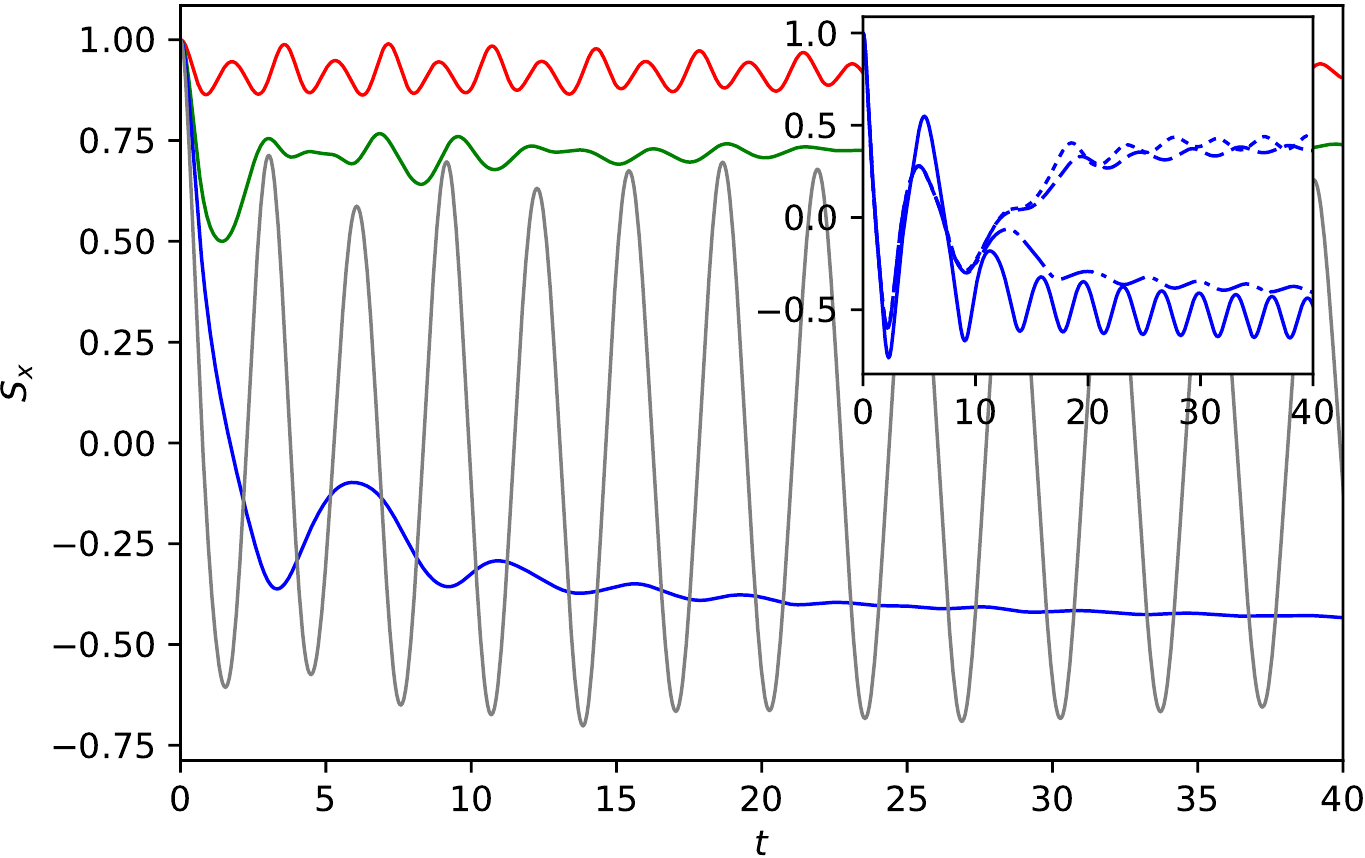}
\caption{%(Color online) 
Evolution of the order parameter $S_x(t)$ for $J=0.67$, $g=0.5$, 0.83, 1, 1.33 (red, green, blue, gray), with $\bar{\lambda}=\lambda+J=1$ and %system size 
$N=400$, as obtained from MPS-TDVP simulations. 
\textit{Inset:} Sensitivity of $S_x(t)$  to
the system size $N$ in the chaotic dynamical ferromagnetic phase, 
for a system with  $J=0.5$, %$\bar{\lambda}=1$, 
$g=1.1$, and bond dimension $D=128$.
The magnetization $S_x$ approaches a positive value for small %system sizes 
$N=123$, 124 (dashed, dotted). However, upon adding just one spin ($N=125$, dash-dotted), $S_x$ reverses its sign and $\bar S_x$ converges to a negative value, which is also observed in larger systems with $N=400$ (solid line). For further details see~\cite{SM}.
%
%Note that the damping in the oscillations of the order parameter is a finite-size effect.
%
%
}
\label{fig:tdvp}
\end{figure}
%%%%%%%%%%%%%%%%%%%%%%%%%%%%%%%%%%%%%%%%%%
%%%%%%%%%%%%%

%Since the system is isolated and energy is conserved,  it is unclear whether $S_x(t)$ would  continue hopping from one well to the other. 
%\sout{Perturbative estimates \cite{SM} (with $J\ll g,\lambda$)  close to the dynamical critical point $g\simeq\bar{\lambda}$ reveal that the quantum feedback $\delta^{\alpha\beta}(t)$ responsible for the transitions between these two wells decays algebraically in time. For instance, $\delta^{qq}(t)$ 
%affecting the dynamics of $\phi(t)$, is characterised by beats of {frequencies $2J$ and $2\bar{\lambda}$, }
%with an amplitude decaying as $\propto (Jt)^{-1/2}$ for  times $J^{-1}\ll t\ll J^{-2}$.
%
%This resembles the inhomogeneous dephasing mechanism responsible for the relaxation of local observables after a quench in isolated quantum many-body systems. Accordingly, the dynamics will be on this time scale described by %just 
%the classical equations with time-independent feedback terms.}% and $S_x$ will eventually localise in one of the two possible ferromagnetic states.

The dynamical phases {discussed here turn out to be robust against the perturbation 
of $J$} %introduced 
even for values %of 
$J \simeq 0.67$ at which the low-density spin-wave expansion need not 
be accurate. To show this, we performed numerical simulations by using a
time-dependent variational principle on the matrix product state manifold
(MPS-TDVP) \cite{LOV15,HLO14}, resulting in the evolution reported in
Fig.~\ref{fig:tdvp}; {this approach allows us to explore the dynamics of
  $S_x(t)$ up to times of the order $\sim 60 \bar{\lambda}^{-1}$.}
For $g\lesssim \bar\lambda$ we find a ferromagnetic region with $\bar S_x \neq 0$ of the same sign as the initial magnetisation $S_x(t=0)$. 
%%
%%%%%%%%%%%%%%%%%%%%%%%%%%%%%
%%
%%
For large values of %the magnetic fields
$g\gtrsim \bar\lambda$, instead, we find a paramagnetic phase with $\bar S_x = 0$ \cite{SM}, 
while for intermediate values, $\bar S_x$ does not vanish but it may have a sign opposite to that of 
$S_x(0)$; this observation is consistent with what observed at smaller values of $J$, see Fig.~\ref{fig:tdvp}.  In addition, in this regime, the final value of $\bar S_x$ may sensibly depend on the system size $N$:
For $N\approx 100$ (see caption of Fig.~\ref{fig:tdvp})
we observe $\bar S_x\neq 0$ of the same sign as $S_x(t=0)$, while for a slightly larger system $N=125$, $S_x(t)$ at long times has the opposite sign, which is eventually observed also in a system with $N=400$, as shown by the inset of Fig.~\ref{fig:tdvp}. 
This is consistent {with the sensitivity %with respect 
to the parameters} predicted by the spin-wave approach, see Fig.~\ref{fig1}.
{These %In conclusion, the 
numerical simulations of the exact %many-body 
quantum evolution fully confirm --- and even extend to a larger region of the phase diagram --- the scenario outlined by the time-dependent spin-wave theory, i.e., the robustness of the two dynamical phases and the emergence of a %onset of a 
chaotic region in between.} 

\emph{Perspectives} --- In summary, %\sout{the dynamical phase diagram of the model Eq.~\eqref{eq:ham} in Fig.~\ref{fig1} demonstrates that} 
the non-equilibrium quantum fluctuations due {to spin-wave excitations modify} % alters
qualitatively the mean-field phase diagram, turning the $J=0$ quantum critical point  into a phase with %novel 
{unusual dynamical properties.} 
%hopping between
%chaotic collective dynamics.
%
%
The non-equilibrium spin-wave theory at the core of this work can be straightforwardly extended to a  wide variety of  spin systems, in higher dimensions, with other types of integrability breaking terms (of short or long-range character) or  non-equilibrium protocols: a chaotic dynamical phase always arises whenever a mean-field system undergoing a %isotropic or 
ferromagnetic transition is subject to the impact of out-of-equilibrium quantum fluctuations~\cite{Long}. %\sout{In these cases, we expect that the scenario outlined in Fig.~\ref{fig1} will remain qualitatively unaltered. }
In addition, the phenomena discussed here could be within experimental reach, considering recent progress in realising spin models~\cite{Simon11,*Meinert13,*Labuhn16} 
{as well as 
%and in view of recent and significant advances
in highlighting universal scaling behaviour close to}  dynamical critical points using cold gases~\cite{Oberthaler2015}.
%
%\sout{We plan to investigate some of these issues in future works.}
%

\emph{Acknowledgements ---}  We thank M. Fabrizio for interesting discussions and A. Rosch for insightful comments on the manuscript.
A.~L. acknowledges hospitality from the University of Cologne.
J.~M. acknowledges support from the Alexander von Humboldt foundation. B.~\v{Z}. is
supported by the ERC project OMNES. 

\bibliography{biblio}

\begin{thebibliography}{45}%
\makeatletter
\providecommand \@ifxundefined [1]{%
 \@ifx{#1\undefined}
}%
\providecommand \@ifnum [1]{%
 \ifnum #1\expandafter \@firstoftwo
 \else \expandafter \@secondoftwo
 \fi
}%
\providecommand \@ifx [1]{%
 \ifx #1\expandafter \@firstoftwo
 \else \expandafter \@secondoftwo
 \fi
}%
\providecommand \natexlab [1]{#1}%
\providecommand \enquote  [1]{``#1''}%
\providecommand \bibnamefont  [1]{#1}%
\providecommand \bibfnamefont [1]{#1}%
\providecommand \citenamefont [1]{#1}%
\providecommand \href@noop [0]{\@secondoftwo}%
\providecommand \href [0]{\begingroup \@sanitize@url \@href}%
\providecommand \@href[1]{\@@startlink{#1}\@@href}%
\providecommand \@@href[1]{\endgroup#1\@@endlink}%
\providecommand \@sanitize@url [0]{\catcode `\\12\catcode `\$12\catcode
  `\&12\catcode `\#12\catcode `\^12\catcode `\_12\catcode `\%12\relax}%
\providecommand \@@startlink[1]{}%
\providecommand \@@endlink[0]{}%
\providecommand \url  [0]{\begingroup\@sanitize@url \@url }%
\providecommand \@url [1]{\endgroup\@href {#1}{\urlprefix }}%
\providecommand \urlprefix  [0]{URL }%
\providecommand \Eprint [0]{\href }%
\providecommand \doibase [0]{http://dx.doi.org/}%
\providecommand \selectlanguage [0]{\@gobble}%
\providecommand \bibinfo  [0]{\@secondoftwo}%
\providecommand \bibfield  [0]{\@secondoftwo}%
\providecommand \translation [1]{[#1]}%
\providecommand \BibitemOpen [0]{}%
\providecommand \bibitemStop [0]{}%
\providecommand \bibitemNoStop [0]{.\EOS\space}%
\providecommand \EOS [0]{\spacefactor3000\relax}%
\providecommand \BibitemShut  [1]{\csname bibitem#1\endcsname}%
\let\auto@bib@innerbib\@empty
%</preamble>
\bibitem [{\citenamefont {Strzalko}\ \emph {et~al.}(2009)\citenamefont
  {Strzalko}, \citenamefont {Grabski}, \citenamefont {Perlikowski},
  \citenamefont {Stefanski},\ and\ \citenamefont {Kapitaniak}}]{Dicebook}%
  \BibitemOpen
  \bibfield  {author} {\bibinfo {author} {\bibfnamefont {J.}~\bibnamefont
  {Strzalko}}, \bibinfo {author} {\bibfnamefont {J.}~\bibnamefont {Grabski}},
  \bibinfo {author} {\bibfnamefont {P.}~\bibnamefont {Perlikowski}}, \bibinfo
  {author} {\bibfnamefont {A.}~\bibnamefont {Stefanski}}, \ and\ \bibinfo
  {author} {\bibfnamefont {T.}~\bibnamefont {Kapitaniak}},\ }\href@noop {}
  {\emph {\bibinfo {title} {{Dynamics of Gambling: Origins of Randomness in
  Mechanical Systems, Lecture Notes in Physics 792}}}}\ (\bibinfo  {publisher}
  {Springer, Berlin},\ \bibinfo {year} {2009})\BibitemShut {NoStop}%
\bibitem [{\citenamefont {Polkovnikov}\ \emph {et~al.}(2011)\citenamefont
  {Polkovnikov}, \citenamefont {Sengupta}, \citenamefont {Silva},\ and\
  \citenamefont {Vengalattore}}]{PolkovnikovRMP}%
  \BibitemOpen
  \bibfield  {author} {\bibinfo {author} {\bibfnamefont {A.}~\bibnamefont
  {Polkovnikov}}, \bibinfo {author} {\bibfnamefont {K.}~\bibnamefont
  {Sengupta}}, \bibinfo {author} {\bibfnamefont {A.}~\bibnamefont {Silva}}, \
  and\ \bibinfo {author} {\bibfnamefont {M.}~\bibnamefont {Vengalattore}},\
  }\href {\doibase 10.1103/RevModPhys.83.863} {\bibfield  {journal} {\bibinfo
  {journal} {Rev. Mod. Phys.}\ }\textbf {\bibinfo {volume} {83}},\ \bibinfo
  {pages} {863} (\bibinfo {year} {2011})}\BibitemShut {NoStop}%
\bibitem [{\citenamefont {Lamacraft}\ and\ \citenamefont
  {Moore}(2012)}]{Lamacraft2012}%
  \BibitemOpen
  \bibfield  {author} {\bibinfo {author} {\bibfnamefont {A.}~\bibnamefont
  {Lamacraft}}\ and\ \bibinfo {author} {\bibfnamefont {J.}~\bibnamefont
  {Moore}},\ }in\ \href@noop {} {\emph {\bibinfo {booktitle} {{Ultracold
  Bosonic and Fermionic Gases}}}},\ \bibinfo {editor} {edited by\ \bibinfo
  {editor} {\bibfnamefont {K.}~\bibnamefont {Levin}}, \bibinfo {editor}
  {\bibfnamefont {A.}~\bibnamefont {Fetter}}, \ and\ \bibinfo {editor}
  {\bibfnamefont {D.}~\bibnamefont {Stamper-Kurn}}}\ (\bibinfo  {publisher}
  {Elsevier, Amsterdam},\ \bibinfo {year} {2012})\ Chap.~\bibinfo {chapter}
  {7}\BibitemShut {NoStop}%
\bibitem [{\citenamefont {Gogolin}\ and\ \citenamefont
  {Eisert}(2016)}]{Eisert2015a}%
  \BibitemOpen
  \bibfield  {author} {\bibinfo {author} {\bibfnamefont {C.}~\bibnamefont
  {Gogolin}}\ and\ \bibinfo {author} {\bibfnamefont {J.}~\bibnamefont
  {Eisert}},\ }\href@noop {} {\bibfield  {journal} {\bibinfo  {journal} {Rep.
  Prog. Phys. 79, 056001}\ } (\bibinfo {year} {2016})}\BibitemShut {NoStop}%
\bibitem [{\citenamefont {Greiner}\ \emph
  {et~al.}(2002{\natexlab{a}})\citenamefont {Greiner}, \citenamefont {Mandel},
  \citenamefont {Hansch},\ and\ \citenamefont {Bloch}}]{Greiner2002a}%
  \BibitemOpen
  \bibfield  {author} {\bibinfo {author} {\bibfnamefont {M.}~\bibnamefont
  {Greiner}}, \bibinfo {author} {\bibfnamefont {O.}~\bibnamefont {Mandel}},
  \bibinfo {author} {\bibfnamefont {T.~W.}\ \bibnamefont {Hansch}}, \ and\
  \bibinfo {author} {\bibfnamefont {I.}~\bibnamefont {Bloch}},\ }\href@noop {}
  {\bibfield  {journal} {\bibinfo  {journal} {Nature}\ }\textbf {\bibinfo
  {volume} {419}},\ \bibinfo {pages} {51} (\bibinfo {year}
  {2002}{\natexlab{a}})}\BibitemShut {NoStop}%
\bibitem [{\citenamefont {Greiner}\ \emph
  {et~al.}(2002{\natexlab{b}})\citenamefont {Greiner}, \citenamefont {Mandel},
  \citenamefont {Esslinger}, \citenamefont {H{\"a}nsch},\ and\ \citenamefont
  {Bloch}}]{Greiner2002b}%
  \BibitemOpen
  \bibfield  {author} {\bibinfo {author} {\bibfnamefont {M.}~\bibnamefont
  {Greiner}}, \bibinfo {author} {\bibfnamefont {O.}~\bibnamefont {Mandel}},
  \bibinfo {author} {\bibfnamefont {T.}~\bibnamefont {Esslinger}}, \bibinfo
  {author} {\bibfnamefont {T.~W.}\ \bibnamefont {H{\"a}nsch}}, \ and\ \bibinfo
  {author} {\bibfnamefont {I.}~\bibnamefont {Bloch}},\ }\href {\doibase
  10.1038/415039a} {\bibfield  {journal} {\bibinfo  {journal} {Nature}\
  }\textbf {\bibinfo {volume} {415}},\ \bibinfo {pages} {39} (\bibinfo {year}
  {2002}{\natexlab{b}})}\BibitemShut {NoStop}%
\bibitem [{\citenamefont {Bloch}\ \emph {et~al.}(2008)\citenamefont {Bloch},
  \citenamefont {Dalibard},\ and\ \citenamefont {Zwerger}}]{Bloch2008}%
  \BibitemOpen
  \bibfield  {author} {\bibinfo {author} {\bibfnamefont {I.}~\bibnamefont
  {Bloch}}, \bibinfo {author} {\bibfnamefont {J.}~\bibnamefont {Dalibard}}, \
  and\ \bibinfo {author} {\bibfnamefont {W.}~\bibnamefont {Zwerger}},\ }\href
  {\doibase 10.1103/RevModPhys.80.885} {\bibfield  {journal} {\bibinfo
  {journal} {Rev. Mod. Phys.}\ }\textbf {\bibinfo {volume} {80}},\ \bibinfo
  {pages} {885} (\bibinfo {year} {2008})}\BibitemShut {NoStop}%
\bibitem [{\citenamefont {Trotzky}\ \emph {et~al.}(2008)\citenamefont
  {Trotzky}, \citenamefont {Cheinet}, \citenamefont {F{\"o}lling},
  \citenamefont {Feld}, \citenamefont {Schnorrberger}, \citenamefont {Rey},
  \citenamefont {Polkovnikov}, \citenamefont {Demler}, \citenamefont {Lukin},\
  and\ \citenamefont {Bloch}}]{Trotzky2008}%
  \BibitemOpen
  \bibfield  {author} {\bibinfo {author} {\bibfnamefont {S.}~\bibnamefont
  {Trotzky}}, \bibinfo {author} {\bibfnamefont {P.}~\bibnamefont {Cheinet}},
  \bibinfo {author} {\bibfnamefont {S.}~\bibnamefont {F{\"o}lling}}, \bibinfo
  {author} {\bibfnamefont {M.}~\bibnamefont {Feld}}, \bibinfo {author}
  {\bibfnamefont {U.}~\bibnamefont {Schnorrberger}}, \bibinfo {author}
  {\bibfnamefont {A.~M.}\ \bibnamefont {Rey}}, \bibinfo {author} {\bibfnamefont
  {A.}~\bibnamefont {Polkovnikov}}, \bibinfo {author} {\bibfnamefont {E.~A.}\
  \bibnamefont {Demler}}, \bibinfo {author} {\bibfnamefont {M.~D.}\
  \bibnamefont {Lukin}}, \ and\ \bibinfo {author} {\bibfnamefont
  {I.}~\bibnamefont {Bloch}},\ }\href {\doibase 10.1126/science.1150841}
  {\bibfield  {journal} {\bibinfo  {journal} {Science}\ }\textbf {\bibinfo
  {volume} {319}},\ \bibinfo {pages} {295} (\bibinfo {year}
  {2008})}\BibitemShut {NoStop}%
\bibitem [{\citenamefont {Cheneau}\ \emph {et~al.}(2012)\citenamefont
  {Cheneau}, \citenamefont {Barmettler}, \citenamefont {Poletti}, \citenamefont
  {Endres}, \citenamefont {Schausz}, \citenamefont {Fukuhara}, \citenamefont
  {Gross}, \citenamefont {Bloch}, \citenamefont {Kollath},\ and\ \citenamefont
  {Kuhr}}]{Cheneau2012}%
  \BibitemOpen
  \bibfield  {author} {\bibinfo {author} {\bibfnamefont {M.}~\bibnamefont
  {Cheneau}}, \bibinfo {author} {\bibfnamefont {P.}~\bibnamefont {Barmettler}},
  \bibinfo {author} {\bibfnamefont {D.}~\bibnamefont {Poletti}}, \bibinfo
  {author} {\bibfnamefont {M.}~\bibnamefont {Endres}}, \bibinfo {author}
  {\bibfnamefont {P.}~\bibnamefont {Schausz}}, \bibinfo {author} {\bibfnamefont
  {T.}~\bibnamefont {Fukuhara}}, \bibinfo {author} {\bibfnamefont
  {C.}~\bibnamefont {Gross}}, \bibinfo {author} {\bibfnamefont
  {I.}~\bibnamefont {Bloch}}, \bibinfo {author} {\bibfnamefont
  {C.}~\bibnamefont {Kollath}}, \ and\ \bibinfo {author} {\bibfnamefont
  {S.}~\bibnamefont {Kuhr}},\ }\href@noop {} {\bibfield  {journal} {\bibinfo
  {journal} {Nature}\ }\textbf {\bibinfo {volume} {481}},\ \bibinfo {pages}
  {484} (\bibinfo {year} {2012})}\BibitemShut {NoStop}%
\bibitem [{\citenamefont {Kaufman}\ and\ \citenamefont {\emph{et
  al}}(2016)}]{Kauf16}%
  \BibitemOpen
  \bibfield  {author} {\bibinfo {author} {\bibfnamefont {A.}~\bibnamefont
  {Kaufman}}\ and\ \bibinfo {author} {\bibnamefont {\emph{et al}}},\
  }\href@noop {} {\bibfield  {journal} {\bibinfo  {journal} {Science}\ }\textbf
  {\bibinfo {volume} {353}},\ \bibinfo {pages} {794} (\bibinfo {year}
  {2016})}\BibitemShut {NoStop}%
\bibitem [{\citenamefont {Schweigler}\ \emph {et~al.}(2017)\citenamefont
  {Schweigler}, \citenamefont {Kasper}, \citenamefont {Erne}, \citenamefont
  {Rauer}, \citenamefont {Langen}, \citenamefont {Gasenzer}, \citenamefont
  {Berges},\ and\ \citenamefont {Schmiedmayer}}]{Jorg15}%
  \BibitemOpen
  \bibfield  {author} {\bibinfo {author} {\bibfnamefont {T.}~\bibnamefont
  {Schweigler}}, \bibinfo {author} {\bibfnamefont {V.}~\bibnamefont {Kasper}},
  \bibinfo {author} {\bibfnamefont {S.}~\bibnamefont {Erne}}, \bibinfo {author}
  {\bibfnamefont {B.}~\bibnamefont {Rauer}}, \bibinfo {author} {\bibfnamefont
  {T.}~\bibnamefont {Langen}}, \bibinfo {author} {\bibfnamefont
  {T.}~\bibnamefont {Gasenzer}}, \bibinfo {author} {\bibfnamefont
  {J.}~\bibnamefont {Berges}}, \ and\ \bibinfo {author} {\bibfnamefont
  {J.}~\bibnamefont {Schmiedmayer}},\ }\href@noop {} {\bibfield  {journal}
  {\bibinfo  {journal} {Nature}\ }\textbf {\bibinfo {volume} {545}},\ \bibinfo
  {pages} {323} (\bibinfo {year} {2017})}\BibitemShut {NoStop}%
\bibitem [{\citenamefont {Rauer}\ \emph {et~al.}(2017)\citenamefont {Rauer},
  \citenamefont {Erne}, \citenamefont {Schweigler}, \citenamefont {Cataldini},
  \citenamefont {Tajik},\ and\ \citenamefont {Schmiedmayer}}]{rauer17}%
  \BibitemOpen
  \bibfield  {author} {\bibinfo {author} {\bibfnamefont {B.}~\bibnamefont
  {Rauer}}, \bibinfo {author} {\bibfnamefont {S.}~\bibnamefont {Erne}},
  \bibinfo {author} {\bibfnamefont {T.}~\bibnamefont {Schweigler}}, \bibinfo
  {author} {\bibfnamefont {F.}~\bibnamefont {Cataldini}}, \bibinfo {author}
  {\bibfnamefont {M.}~\bibnamefont {Tajik}}, \ and\ \bibinfo {author}
  {\bibfnamefont {J.}~\bibnamefont {Schmiedmayer}},\ }\href@noop {} {\bibfield
  {journal} {\bibinfo  {journal} {arXiv:1705.08231}\ } (\bibinfo {year}
  {2017})}\BibitemShut {NoStop}%
\bibitem [{\citenamefont {Hess}\ \emph {et~al.}(2017)\citenamefont {Hess},
  \citenamefont {Becker}, \citenamefont {Kaplan}, \citenamefont {Kyprianidis},
  \citenamefont {Lee}, \citenamefont {Neyenhuis}, \citenamefont {Pagano},
  \citenamefont {Richerme}, \citenamefont {Senko}, \citenamefont {Smith},
  \citenamefont {Tan}, \citenamefont {Zhang},\ and\ \citenamefont
  {Monroe}}]{hess17}%
  \BibitemOpen
  \bibfield  {author} {\bibinfo {author} {\bibfnamefont {P.~W.}\ \bibnamefont
  {Hess}}, \bibinfo {author} {\bibfnamefont {P.}~\bibnamefont {Becker}},
  \bibinfo {author} {\bibfnamefont {H.~B.}\ \bibnamefont {Kaplan}}, \bibinfo
  {author} {\bibfnamefont {A.}~\bibnamefont {Kyprianidis}}, \bibinfo {author}
  {\bibfnamefont {A.~C.}\ \bibnamefont {Lee}}, \bibinfo {author} {\bibfnamefont
  {B.}~\bibnamefont {Neyenhuis}}, \bibinfo {author} {\bibfnamefont
  {G.}~\bibnamefont {Pagano}}, \bibinfo {author} {\bibfnamefont
  {P.}~\bibnamefont {Richerme}}, \bibinfo {author} {\bibfnamefont
  {C.}~\bibnamefont {Senko}}, \bibinfo {author} {\bibfnamefont
  {J.}~\bibnamefont {Smith}}, \bibinfo {author} {\bibfnamefont {W.~L.}\
  \bibnamefont {Tan}}, \bibinfo {author} {\bibfnamefont {J.}~\bibnamefont
  {Zhang}}, \ and\ \bibinfo {author} {\bibfnamefont {C.}~\bibnamefont
  {Monroe}},\ }\href@noop {} {\bibfield  {journal} {\bibinfo  {journal}
  {arXiv:1704.02439}\ } (\bibinfo {year} {2017})}\BibitemShut {NoStop}%
\bibitem [{\citenamefont {Sciolla}\ and\ \citenamefont
  {Biroli}(2011)}]{Sciolla2011}%
  \BibitemOpen
  \bibfield  {author} {\bibinfo {author} {\bibfnamefont {B.}~\bibnamefont
  {Sciolla}}\ and\ \bibinfo {author} {\bibfnamefont {G.}~\bibnamefont
  {Biroli}},\ }\href@noop {} {\bibfield  {journal} {\bibinfo  {journal} {J.
  Stat. Mech.}\ ,\ \bibinfo {pages} {P11003}} (\bibinfo {year}
  {2011})}\BibitemShut {NoStop}%
\bibitem [{\citenamefont {Smacchia}\ \emph {et~al.}(2015)\citenamefont
  {Smacchia}, \citenamefont {Knap}, \citenamefont {Demler},\ and\ \citenamefont
  {Silva}}]{Smacchia2014}%
  \BibitemOpen
  \bibfield  {author} {\bibinfo {author} {\bibfnamefont {P.}~\bibnamefont
  {Smacchia}}, \bibinfo {author} {\bibfnamefont {M.}~\bibnamefont {Knap}},
  \bibinfo {author} {\bibfnamefont {E.}~\bibnamefont {Demler}}, \ and\ \bibinfo
  {author} {\bibfnamefont {A.}~\bibnamefont {Silva}},\ }\href {\doibase
  10.1103/PhysRevB.91.205136} {\bibfield  {journal} {\bibinfo  {journal} {Phys.
  Rev. B}\ }\textbf {\bibinfo {volume} {91}},\ \bibinfo {pages} {205136}
  (\bibinfo {year} {2015})}\BibitemShut {NoStop}%
\bibitem [{\citenamefont {Maraga}\ \emph {et~al.}(2016)\citenamefont {Maraga},
  \citenamefont {Smacchia},\ and\ \citenamefont {Silva}}]{Maraga2016}%
  \BibitemOpen
  \bibfield  {author} {\bibinfo {author} {\bibfnamefont {A.}~\bibnamefont
  {Maraga}}, \bibinfo {author} {\bibfnamefont {P.}~\bibnamefont {Smacchia}}, \
  and\ \bibinfo {author} {\bibfnamefont {A.}~\bibnamefont {Silva}},\ }\href
  {\doibase 10.1103/PhysRevB.94.245122} {\bibfield  {journal} {\bibinfo
  {journal} {Phys. Rev. B}\ }\textbf {\bibinfo {volume} {94}},\ \bibinfo
  {pages} {245122} (\bibinfo {year} {2016})}\BibitemShut {NoStop}%
\bibitem [{\citenamefont {Halimeh}\ \emph {et~al.}(2017)\citenamefont
  {Halimeh}, \citenamefont {Zauner-Stauber}, \citenamefont {McCulloch},
  \citenamefont {de~Vega}, \citenamefont {Schollw\"ock},\ and\ \citenamefont
  {Kastner}}]{Halimeh2016}%
  \BibitemOpen
  \bibfield  {author} {\bibinfo {author} {\bibfnamefont {J.~C.}\ \bibnamefont
  {Halimeh}}, \bibinfo {author} {\bibfnamefont {V.}~\bibnamefont
  {Zauner-Stauber}}, \bibinfo {author} {\bibfnamefont {I.~P.}\ \bibnamefont
  {McCulloch}}, \bibinfo {author} {\bibfnamefont {I.}~\bibnamefont {de~Vega}},
  \bibinfo {author} {\bibfnamefont {U.}~\bibnamefont {Schollw\"ock}}, \ and\
  \bibinfo {author} {\bibfnamefont {M.}~\bibnamefont {Kastner}},\ }\href
  {\doibase 10.1103/PhysRevB.95.024302} {\bibfield  {journal} {\bibinfo
  {journal} {Phys. Rev. B}\ }\textbf {\bibinfo {volume} {95}},\ \bibinfo
  {pages} {024302} (\bibinfo {year} {2017})}\BibitemShut {NoStop}%
\bibitem [{\citenamefont {Homrighausen}\ \emph {et~al.}(2017)\citenamefont
  {Homrighausen}, \citenamefont {Abeling}, \citenamefont {Zauner-Stauber},\
  and\ \citenamefont {Halimeh}}]{Halimeh2017}%
  \BibitemOpen
  \bibfield  {author} {\bibinfo {author} {\bibfnamefont {I.}~\bibnamefont
  {Homrighausen}}, \bibinfo {author} {\bibfnamefont {N.~O.}\ \bibnamefont
  {Abeling}}, \bibinfo {author} {\bibfnamefont {V.}~\bibnamefont
  {Zauner-Stauber}}, \ and\ \bibinfo {author} {\bibfnamefont {J.~C.}\
  \bibnamefont {Halimeh}},\ }\href@noop {} {\bibfield  {journal} {\bibinfo
  {journal} {arXiv:1703.09195}\ } (\bibinfo {year} {2017})}\BibitemShut
  {NoStop}%
\bibitem [{\citenamefont {Gambassi}\ and\ \citenamefont
  {Calabrese}(2011)}]{Gambassi2011}%
  \BibitemOpen
  \bibfield  {author} {\bibinfo {author} {\bibfnamefont {A.}~\bibnamefont
  {Gambassi}}\ and\ \bibinfo {author} {\bibfnamefont {P.}~\bibnamefont
  {Calabrese}},\ }\href {http://stacks.iop.org/0295-5075/95/i=6/a=66007}
  {\bibfield  {journal} {\bibinfo  {journal} {Europhys. Lett.}\ }\textbf
  {\bibinfo {volume} {95}},\ \bibinfo {pages} {66007} (\bibinfo {year}
  {2011})}\BibitemShut {NoStop}%
\bibitem [{\citenamefont {Sciolla}\ and\ \citenamefont
  {Biroli}(2013)}]{Sciolla2013}%
  \BibitemOpen
  \bibfield  {author} {\bibinfo {author} {\bibfnamefont {B.}~\bibnamefont
  {Sciolla}}\ and\ \bibinfo {author} {\bibfnamefont {G.}~\bibnamefont
  {Biroli}},\ }\href {\doibase 10.1103/PhysRevB.88.201110} {\bibfield
  {journal} {\bibinfo  {journal} {Phys. Rev. B}\ }\textbf {\bibinfo {volume}
  {88}},\ \bibinfo {pages} {201110} (\bibinfo {year} {2013})}\BibitemShut
  {NoStop}%
\bibitem [{\citenamefont {Chiocchetta}\ \emph {et~al.}(2015)\citenamefont
  {Chiocchetta}, \citenamefont {Tavora}, \citenamefont {Gambassi},\ and\
  \citenamefont {Mitra}}]{Chiocchetta2015}%
  \BibitemOpen
  \bibfield  {author} {\bibinfo {author} {\bibfnamefont {A.}~\bibnamefont
  {Chiocchetta}}, \bibinfo {author} {\bibfnamefont {M.}~\bibnamefont {Tavora}},
  \bibinfo {author} {\bibfnamefont {A.}~\bibnamefont {Gambassi}}, \ and\
  \bibinfo {author} {\bibfnamefont {A.}~\bibnamefont {Mitra}},\ }\href
  {\doibase 10.1103/PhysRevB.91.220302} {\bibfield  {journal} {\bibinfo
  {journal} {Phys. Rev. B}\ }\textbf {\bibinfo {volume} {91}},\ \bibinfo
  {pages} {220302} (\bibinfo {year} {2015})}\BibitemShut {NoStop}%
\bibitem [{\citenamefont {Heyl}\ \emph {et~al.}(2013)\citenamefont {Heyl},
  \citenamefont {Polkovnikov},\ and\ \citenamefont {Kehrein}}]{Heyl2013}%
  \BibitemOpen
  \bibfield  {author} {\bibinfo {author} {\bibfnamefont {M.}~\bibnamefont
  {Heyl}}, \bibinfo {author} {\bibfnamefont {A.}~\bibnamefont {Polkovnikov}}, \
  and\ \bibinfo {author} {\bibfnamefont {S.}~\bibnamefont {Kehrein}},\ }\href
  {\doibase 10.1103/PhysRevLett.110.135704} {\bibfield  {journal} {\bibinfo
  {journal} {Phys. Rev. Lett.}\ }\textbf {\bibinfo {volume} {110}},\ \bibinfo
  {pages} {135704} (\bibinfo {year} {2013})}\BibitemShut {NoStop}%
\bibitem [{\citenamefont {{Zunkovic}}\ \emph {et~al.}(2016)\citenamefont
  {{Zunkovic}}, \citenamefont {{Heyl}}, \citenamefont {{Knap}},\ and\
  \citenamefont {{Silva}}}]{Bojan2016b}%
  \BibitemOpen
  \bibfield  {author} {\bibinfo {author} {\bibfnamefont {B.}~\bibnamefont
  {{Zunkovic}}}, \bibinfo {author} {\bibfnamefont {M.}~\bibnamefont {{Heyl}}},
  \bibinfo {author} {\bibfnamefont {M.}~\bibnamefont {{Knap}}}, \ and\ \bibinfo
  {author} {\bibfnamefont {A.}~\bibnamefont {{Silva}}},\ }\href@noop {}
  {\bibfield  {journal} {\bibinfo  {journal} {ArXiv e-prints}\ } (\bibinfo
  {year} {2016})},\ \Eprint {http://arxiv.org/abs/1609.08482}
  {arXiv:1609.08482} \BibitemShut {NoStop}%
\bibitem [{\citenamefont {Chandran}\ \emph {et~al.}(2013)\citenamefont
  {Chandran}, \citenamefont {Nanduri}, \citenamefont {Gubser},\ and\
  \citenamefont {Sondhi}}]{Chandran2013}%
  \BibitemOpen
  \bibfield  {author} {\bibinfo {author} {\bibfnamefont {A.}~\bibnamefont
  {Chandran}}, \bibinfo {author} {\bibfnamefont {A.}~\bibnamefont {Nanduri}},
  \bibinfo {author} {\bibfnamefont {S.~S.}\ \bibnamefont {Gubser}}, \ and\
  \bibinfo {author} {\bibfnamefont {S.~L.}\ \bibnamefont {Sondhi}},\ }\href
  {\doibase 10.1103/PhysRevB.88.024306} {\bibfield  {journal} {\bibinfo
  {journal} {Phys. Rev. B}\ }\textbf {\bibinfo {volume} {88}},\ \bibinfo
  {pages} {024306} (\bibinfo {year} {2013})}\BibitemShut {NoStop}%
\bibitem [{\citenamefont {Chiocchetta}\ \emph {et~al.}(2017)\citenamefont
  {Chiocchetta}, \citenamefont {Gambassi}, \citenamefont {Diehl},\ and\
  \citenamefont {Marino}}]{Chiocchetta2017}%
  \BibitemOpen
  \bibfield  {author} {\bibinfo {author} {\bibfnamefont {A.}~\bibnamefont
  {Chiocchetta}}, \bibinfo {author} {\bibfnamefont {A.}~\bibnamefont
  {Gambassi}}, \bibinfo {author} {\bibfnamefont {S.}~\bibnamefont {Diehl}}, \
  and\ \bibinfo {author} {\bibfnamefont {J.}~\bibnamefont {Marino}},\ }\href
  {\doibase 10.1103/PhysRevLett.118.135701} {\bibfield  {journal} {\bibinfo
  {journal} {Phys. Rev. Lett.}\ }\textbf {\bibinfo {volume} {118}},\ \bibinfo
  {pages} {135701} (\bibinfo {year} {2017})}\BibitemShut {NoStop}%
\bibitem [{\citenamefont {Maraga}\ \emph {et~al.}(2015)\citenamefont {Maraga},
  \citenamefont {Chiocchetta}, \citenamefont {Mitra},\ and\ \citenamefont
  {Gambassi}}]{Maraga2015}%
  \BibitemOpen
  \bibfield  {author} {\bibinfo {author} {\bibfnamefont {A.}~\bibnamefont
  {Maraga}}, \bibinfo {author} {\bibfnamefont {A.}~\bibnamefont {Chiocchetta}},
  \bibinfo {author} {\bibfnamefont {A.}~\bibnamefont {Mitra}}, \ and\ \bibinfo
  {author} {\bibfnamefont {A.}~\bibnamefont {Gambassi}},\ }\href {\doibase
  10.1103/PhysRevE.92.042151} {\bibfield  {journal} {\bibinfo  {journal} {Phys.
  Rev. E}\ }\textbf {\bibinfo {volume} {92}},\ \bibinfo {pages} {042151}
  (\bibinfo {year} {2015})}\BibitemShut {NoStop}%
\bibitem [{\citenamefont {Janssen}\ \emph {et~al.}(1989)\citenamefont
  {Janssen}, \citenamefont {Schaub},\ and\ \citenamefont
  {Schmittmann}}]{Janssen1989}%
  \BibitemOpen
  \bibfield  {author} {\bibinfo {author} {\bibfnamefont {H.-K.}\ \bibnamefont
  {Janssen}}, \bibinfo {author} {\bibfnamefont {B.}~\bibnamefont {Schaub}}, \
  and\ \bibinfo {author} {\bibfnamefont {B.}~\bibnamefont {Schmittmann}},\
  }\href {\doibase 10.1007/BF01319383} {\bibfield  {journal} {\bibinfo
  {journal} {Z. Phys. B Cond. Mat.}\ }\textbf {\bibinfo {volume} {73}},\
  \bibinfo {pages} {539} (\bibinfo {year} {1989})}\BibitemShut {NoStop}%
\bibitem [{\citenamefont {Calabrese}\ and\ \citenamefont
  {Gambassi}(2005)}]{Calabrese2005}%
  \BibitemOpen
  \bibfield  {author} {\bibinfo {author} {\bibfnamefont {P.}~\bibnamefont
  {Calabrese}}\ and\ \bibinfo {author} {\bibfnamefont {A.}~\bibnamefont
  {Gambassi}},\ }\href {http://stacks.iop.org/0305-4470/38/i=18/a=R01}
  {\bibfield  {journal} {\bibinfo  {journal} {J. Phys. A: Math. Gen.}\ }\textbf
  {\bibinfo {volume} {38}},\ \bibinfo {pages} {R133} (\bibinfo {year}
  {2005})}\BibitemShut {NoStop}%
\bibitem [{\citenamefont {Schir{\'o}}\ and\ \citenamefont
  {Fabrizio}(2011)}]{Schiro2011}%
  \BibitemOpen
  \bibfield  {author} {\bibinfo {author} {\bibfnamefont {M.}~\bibnamefont
  {Schir{\'o}}}\ and\ \bibinfo {author} {\bibfnamefont {M.}~\bibnamefont
  {Fabrizio}},\ }\href {\doibase 10.1103/PhysRevB.83.165105} {\bibfield
  {journal} {\bibinfo  {journal} {Phys. Rev. B}\ }\textbf {\bibinfo {volume}
  {83}},\ \bibinfo {pages} {165105} (\bibinfo {year} {2011})}\BibitemShut
  {NoStop}%
\bibitem [{\citenamefont {Sandri}\ \emph {et~al.}(2012)\citenamefont {Sandri},
  \citenamefont {Schiro},\ and\ \citenamefont {Fabrizio}}]{Sandri12}%
  \BibitemOpen
  \bibfield  {author} {\bibinfo {author} {\bibfnamefont {M.}~\bibnamefont
  {Sandri}}, \bibinfo {author} {\bibfnamefont {M.~J.}\ \bibnamefont {Schiro}},
  \ and\ \bibinfo {author} {\bibfnamefont {M.}~\bibnamefont {Fabrizio}},\
  }\href@noop {} {\bibfield  {journal} {\bibinfo  {journal} {Phys. Rev. B.}\
  }\textbf {\bibinfo {volume} {86}},\ \bibinfo {pages} {075122} (\bibinfo
  {year} {2012})}\BibitemShut {NoStop}%
\bibitem [{\citenamefont {Peronaci}\ \emph {et~al.}(2015)\citenamefont
  {Peronaci}, \citenamefont {Schir{\'o}},\ and\ \citenamefont
  {Capone}}]{Peronaci2015}%
  \BibitemOpen
  \bibfield  {author} {\bibinfo {author} {\bibfnamefont {F.}~\bibnamefont
  {Peronaci}}, \bibinfo {author} {\bibfnamefont {M.}~\bibnamefont
  {Schir{\'o}}}, \ and\ \bibinfo {author} {\bibfnamefont {M.}~\bibnamefont
  {Capone}},\ }\href {\doibase 10.1103/PhysRevLett.115.257001} {\bibfield
  {journal} {\bibinfo  {journal} {Phys. Rev. Lett.}\ }\textbf {\bibinfo
  {volume} {115}},\ \bibinfo {pages} {257001} (\bibinfo {year}
  {2015})}\BibitemShut {NoStop}%
\bibitem [{\citenamefont {Lipkin}\ \emph {et~al.}()\citenamefont {Lipkin},
  \citenamefont {Meshkov},\ and\ \citenamefont {Glick}}]{LMG}%
  \BibitemOpen
  \bibfield  {author} {\bibinfo {author} {\bibfnamefont {H.}~\bibnamefont
  {Lipkin}}, \bibinfo {author} {\bibfnamefont {N.}~\bibnamefont {Meshkov}}, \
  and\ \bibinfo {author} {\bibfnamefont {A.}~\bibnamefont {Glick}},\
  }\href@noop {} {\bibinfo  {journal} {Nucl. Phys., 62, 188 (1965)}\
  }\BibitemShut {NoStop}%
\bibitem [{\citenamefont {Dutta}\ and\ \citenamefont
  {Bhattacharjee}(2001)}]{dutta01}%
  \BibitemOpen
\bibfield  {journal} {  }\bibfield  {author} {\bibinfo {author} {\bibfnamefont
  {A.}~\bibnamefont {Dutta}}\ and\ \bibinfo {author} {\bibfnamefont {J.~K.}\
  \bibnamefont {Bhattacharjee}},\ }\href {\doibase 10.1103/PhysRevB.64.184106}
  {\bibfield  {journal} {\bibinfo  {journal} {Phys. Rev. B}\ }\textbf {\bibinfo
  {volume} {64}},\ \bibinfo {pages} {184106} (\bibinfo {year}
  {2001})}\BibitemShut {NoStop}%
\bibitem [{\citenamefont {Knap}\ \emph {et~al.}(2013)\citenamefont {Knap},
  \citenamefont {Kantian}, \citenamefont {Giamarchi}, \citenamefont {Bloch},
  \citenamefont {Lukin},\ and\ \citenamefont {Demler}}]{knap13}%
  \BibitemOpen
  \bibfield  {author} {\bibinfo {author} {\bibfnamefont {M.}~\bibnamefont
  {Knap}}, \bibinfo {author} {\bibfnamefont {A.}~\bibnamefont {Kantian}},
  \bibinfo {author} {\bibfnamefont {T.}~\bibnamefont {Giamarchi}}, \bibinfo
  {author} {\bibfnamefont {I.}~\bibnamefont {Bloch}}, \bibinfo {author}
  {\bibfnamefont {M.~D.}\ \bibnamefont {Lukin}}, \ and\ \bibinfo {author}
  {\bibfnamefont {E.}~\bibnamefont {Demler}},\ }\href {\doibase
  10.1103/PhysRevLett.111.147205} {\bibfield  {journal} {\bibinfo  {journal}
  {Phys. Rev. Lett.}\ }\textbf {\bibinfo {volume} {111}},\ \bibinfo {pages}
  {147205} (\bibinfo {year} {2013})}\BibitemShut {NoStop}%
\bibitem [{\citenamefont {Zunkovic}\ \emph {et~al.}(2016)\citenamefont
  {Zunkovic}, \citenamefont {Silva},\ and\ \citenamefont {Fabrizio}}]{BoyanMF}%
  \BibitemOpen
  \bibfield  {author} {\bibinfo {author} {\bibfnamefont {B.}~\bibnamefont
  {Zunkovic}}, \bibinfo {author} {\bibfnamefont {A.}~\bibnamefont {Silva}}, \
  and\ \bibinfo {author} {\bibfnamefont {M.}~\bibnamefont {Fabrizio}},\
  }\href@noop {} {\bibfield  {journal} {\bibinfo  {journal} {Phil. Trans. R.
  Soc.}\ }\textbf {\bibinfo {volume} {A 374}},\ \bibinfo {pages} {20150160}
  (\bibinfo {year} {2016})}\BibitemShut {NoStop}%
\bibitem [{\citenamefont {Langen}\ \emph {et~al.}(2016)\citenamefont {Langen},
  \citenamefont {Gasenzer},\ and\ \citenamefont {Schmiedmayer}}]{Langen2016}%
  \BibitemOpen
  \bibfield  {author} {\bibinfo {author} {\bibfnamefont {T.}~\bibnamefont
  {Langen}}, \bibinfo {author} {\bibfnamefont {T.}~\bibnamefont {Gasenzer}}, \
  and\ \bibinfo {author} {\bibfnamefont {J.}~\bibnamefont {Schmiedmayer}},\
  }\href {http://stacks.iop.org/1742-5468/2016/i=6/a=064009} {\bibfield
  {journal} {\bibinfo  {journal} {J. Stat. Mech.}\ ,\ \bibinfo {pages}
  {064009}} (\bibinfo {year} {2016})}\BibitemShut {NoStop}%
\bibitem [{SM()}]{SM}%
  \BibitemOpen
  \href@noop {} {}\bibinfo {note} {See Supplemental Material which contains the
  reference: A. R\"{u}ckriegel, A. Kreisel, and P. Kopietz, Phys. Rev. B 85,
  054422 (2012).}\BibitemShut {Stop}%
\bibitem [{\citenamefont {Lubich}\ \emph {et~al.}(2015)\citenamefont {Lubich},
  \citenamefont {Oseledets},\ and\ \citenamefont {Vandereycken}}]{LOV15}%
  \BibitemOpen
  \bibfield  {author} {\bibinfo {author} {\bibfnamefont {C.}~\bibnamefont
  {Lubich}}, \bibinfo {author} {\bibfnamefont {I.~V.}\ \bibnamefont
  {Oseledets}}, \ and\ \bibinfo {author} {\bibfnamefont {B.}~\bibnamefont
  {Vandereycken}},\ }\href@noop {} {\bibfield  {journal} {\bibinfo  {journal}
  {SIAM J. Numer. Anal}\ }\textbf {\bibinfo {volume} {53}},\ \bibinfo {pages}
  {917} (\bibinfo {year} {2015})}\BibitemShut {NoStop}%
\bibitem [{\citenamefont {Haegeman}\ \emph {et~al.}(2016)\citenamefont
  {Haegeman}, \citenamefont {Lubich}, \citenamefont {Oseledets}, \citenamefont
  {Vandereycken},\ and\ \citenamefont {Verstraete}}]{HLO14}%
  \BibitemOpen
  \bibfield  {author} {\bibinfo {author} {\bibfnamefont {J.}~\bibnamefont
  {Haegeman}}, \bibinfo {author} {\bibfnamefont {C.}~\bibnamefont {Lubich}},
  \bibinfo {author} {\bibfnamefont {I.}~\bibnamefont {Oseledets}}, \bibinfo
  {author} {\bibfnamefont {B.}~\bibnamefont {Vandereycken}}, \ and\ \bibinfo
  {author} {\bibfnamefont {F.}~\bibnamefont {Verstraete}},\ }\href@noop {}
  {\bibfield  {journal} {\bibinfo  {journal} {Phys. Rev. B}\ }\textbf {\bibinfo
  {volume} {94}},\ \bibinfo {pages} {165116} (\bibinfo {year}
  {2016})}\BibitemShut {NoStop}%
\bibitem [{\citenamefont {Sachdev}(2011)}]{Sachdevbook}%
  \BibitemOpen
  \bibfield  {author} {\bibinfo {author} {\bibfnamefont {S.}~\bibnamefont
  {Sachdev}},\ }\href@noop {} {\emph {\bibinfo {title} {{Quantum Phase
  Transitions}}}}\ (\bibinfo  {publisher} {Cambridge University Press},\
  \bibinfo {year} {2011})\BibitemShut {NoStop}%
\bibitem [{\citenamefont {Lerose}\ \emph {et~al.}(2017)\citenamefont {Lerose},
  \citenamefont {Zunkovic}, \citenamefont {Gambassi}, \citenamefont {Marino},\
  and\ \citenamefont {Silva}}]{Long}%
  \BibitemOpen
  \bibfield  {author} {\bibinfo {author} {\bibfnamefont {A.}~\bibnamefont
  {Lerose}}, \bibinfo {author} {\bibfnamefont {B.}~\bibnamefont {Zunkovic}},
  \bibinfo {author} {\bibfnamefont {A.}~\bibnamefont {Gambassi}}, \bibinfo
  {author} {\bibfnamefont {J.}~\bibnamefont {Marino}}, \ and\ \bibinfo {author}
  {\bibfnamefont {A.}~\bibnamefont {Silva}},\ }\href@noop {} {\bibfield
  {journal} {\bibinfo  {journal} {in preparation}\ } (\bibinfo {year}
  {2017})}\BibitemShut {NoStop}%
\bibitem [{\citenamefont {Simon}\ and\ \citenamefont {{\emph{et
  al.}}}(2011)}]{Simon11}%
  \BibitemOpen
  \bibfield  {author} {\bibinfo {author} {\bibfnamefont {J.}~\bibnamefont
  {Simon}}\ and\ \bibinfo {author} {\bibnamefont {{\emph{et al.}}}},\
  }\href@noop {} {\bibfield  {journal} {\bibinfo  {journal} {Nature}\ }\textbf
  {\bibinfo {volume} {472}},\ \bibinfo {pages} {307} (\bibinfo {year}
  {2011})}\BibitemShut {NoStop}%
\bibitem [{\citenamefont {Meinert}\ \emph {et~al.}(2013)\citenamefont
  {Meinert}, \citenamefont {Mark}, \citenamefont {Kirilov}, \citenamefont
  {Lauber}, \citenamefont {Weinmann}, \citenamefont {Daley},\ and\
  \citenamefont {N{\"a}gerl}}]{Meinert13}%
  \BibitemOpen
  \bibfield  {author} {\bibinfo {author} {\bibfnamefont {F.}~\bibnamefont
  {Meinert}}, \bibinfo {author} {\bibfnamefont {M.~J.}\ \bibnamefont {Mark}},
  \bibinfo {author} {\bibfnamefont {E.}~\bibnamefont {Kirilov}}, \bibinfo
  {author} {\bibfnamefont {K.}~\bibnamefont {Lauber}}, \bibinfo {author}
  {\bibfnamefont {P.}~\bibnamefont {Weinmann}}, \bibinfo {author}
  {\bibfnamefont {A.~J.}\ \bibnamefont {Daley}}, \ and\ \bibinfo {author}
  {\bibfnamefont {H.-C.}\ \bibnamefont {N{\"a}gerl}},\ }\href {\doibase
  10.1103/PhysRevLett.111.053003} {\bibfield  {journal} {\bibinfo  {journal}
  {Phys. Rev. Lett.}\ }\textbf {\bibinfo {volume} {111}},\ \bibinfo {pages}
  {053003} (\bibinfo {year} {2013})}\BibitemShut {NoStop}%
\bibitem [{\citenamefont {Labuhn}\ and\ \citenamefont {{\emph{et
  al.}}}(2016)}]{Labuhn16}%
  \BibitemOpen
  \bibfield  {author} {\bibinfo {author} {\bibfnamefont {H.}~\bibnamefont
  {Labuhn}}\ and\ \bibinfo {author} {\bibnamefont {{\emph{et al.}}}},\
  }\href@noop {} {\bibfield  {journal} {\bibinfo  {journal} {Nature}\ }\textbf
  {\bibinfo {volume} {534}},\ \bibinfo {pages} {667} (\bibinfo {year}
  {2016})}\BibitemShut {NoStop}%
\bibitem [{\citenamefont {Nicklas}\ \emph {et~al.}(2015)\citenamefont
  {Nicklas}, \citenamefont {Karl}, \citenamefont {H{\"o}fer}, \citenamefont
  {Johnson}, \citenamefont {Muessel}, \citenamefont {Strobel}, \citenamefont
  {{Tomkovi\ifmmode \check{c}\else \v{c}\fi{}}}, \citenamefont {Gasenzer},\
  and\ \citenamefont {Oberthaler}}]{Oberthaler2015}%
  \BibitemOpen
  \bibfield  {author} {\bibinfo {author} {\bibfnamefont {E.}~\bibnamefont
  {Nicklas}}, \bibinfo {author} {\bibfnamefont {M.}~\bibnamefont {Karl}},
  \bibinfo {author} {\bibfnamefont {M.}~\bibnamefont {H{\"o}fer}}, \bibinfo
  {author} {\bibfnamefont {A.}~\bibnamefont {Johnson}}, \bibinfo {author}
  {\bibfnamefont {W.}~\bibnamefont {Muessel}}, \bibinfo {author} {\bibfnamefont
  {H.}~\bibnamefont {Strobel}}, \bibinfo {author} {\bibfnamefont
  {J.}~\bibnamefont {{Tomkovi\ifmmode \check{c}\else \v{c}\fi{}}}}, \bibinfo
  {author} {\bibfnamefont {T.}~\bibnamefont {Gasenzer}}, \ and\ \bibinfo
  {author} {\bibfnamefont {M.~K.}\ \bibnamefont {Oberthaler}},\ }\href
  {\doibase 10.1103/PhysRevLett.115.245301} {\bibfield  {journal} {\bibinfo
  {journal} {Phys. Rev. Lett.}\ }\textbf {\bibinfo {volume} {115}},\ \bibinfo
  {pages} {245301} (\bibinfo {year} {2015})}\BibitemShut {NoStop}%
\end{thebibliography}%

\newpage
\begin{widetext}

\appendix

\section {\Large Supplemental Material: \\ {Chaotic dynamical ferromagnetic phase \\induced by non-equilibrium quantum 
fluctuations}}

In Section 1, a comprehensive exposition of the time-dependent spin-wave theory is provided. In Section 2, perturbative estimates for the quantum feedback terms near the mean-field dynamical critical point are presented. Finally, in Section 3, details about the numerical study of the full many-body non-equilibrium quantum evolution via TDVP based on matrix product states are discussed.

\section{\large 1. Time-dependent spin-wave theory}
\label{alessio}

As outlined in the Letter, our strategy to study the effect of quantum fluctuations on the dynamics of the spin chain under consideration is based on expressing the quantum dynamical evolution in a time-dependent rotating reference frame ${\mathcal R}$ adapted to the instantaneous direction of the average total spin (see [A. Ruckriegel, \emph{et al.}, Phys. Rev. B 85, 054422 (2012)] for a similar approach).
In ${\mathcal R}$ we can introduce the canonical coordinates corresponding to the spatial fluctuations of the spin field, i.e., to the \emph{spin-waves}, and write the coupled equations of motion for both the total spin and the spin-waves. 
The evolution of the latter enters as a feedback into the  dynamics of the former while the motion of the reference frame is determined self-consistently under the condition that it remains constantly aligned with the instantaneous average value of the total spin. Below we describe the main technical steps of this approach.

\subsection{Time-dependent coordinates}

As usual when dealing with spin-wave theory, it is convenient to consider a spin $\vec{\Sigma}$ of generic value $s$, instead of focussing directly on $s=1/2$, in order to keep track of the classical limit $s\to\infty$. 
Accordingly, we consider the original Hamiltonian (see Eq.~(1) of the Letter), in which $\vec{\sigma}_i \equiv \vec{\Sigma}_i/s$ 
are now normalized spin-$s$ operators: they reduce to the standard Pauli matrices for $s=1/2$, 
%\begin{equation}
%\label{eq:Hs}
%\frac{H}{2s} =  - \frac{2g}{2s}  \sum_i S_i^3  - \frac{4\lambda}{(2s)^2N} \sum_{i,j} S_i^1 S_j^1  -\frac{4J}{(2s)^2} \sum_i S_i^1 S_{i+1}^1,
%\end{equation}
%where, for later convenience, we denote the Cartesian components of the spins by 1, 2, 3 instead of $x$, $y$, $z$, and we define $\vec S_i = (S_i^1,S_i^2,S_i^3)$.
%
while their $1/s$ normalization makes the classical limit $s\to\infty$ well-defined, assuming that time is measured in units of $2s$.

Assuming periodic boundary conditions along the chain, we introduce Fourier modes $\tilde{\vec{\sigma}}_{k}\equiv  \sum_j e^{-ikj} \vec{\sigma}_j$, with wavevectors $k=2\pi n/N$ and $n=0,1,\dots,N-1$; accordingly, the Hamiltonian in Eq.~(1) of the Letter becomes 
\begin{equation}
\label{eq:Hs_fourierbar}
H = 
- g \, \tilde{\sigma}_{0}^z
- \frac{\bar{\lambda}}{N} \left(\tilde{\sigma}_{0}^x\right)^2
- \frac{J}{N } \sum_{k\ne0} \cos k \; \tilde{\sigma}_{k}^{x} \tilde{\sigma}_{-k}^{x},
\end{equation}
where $\bar{\lambda}\equiv\lambda+J$ is the effective value of $\lambda$ which has been renormalized by the term of the Hamiltonian corresponding to the Fourier mode $k=0$. Note that $ \langle \tilde{\vec{\sigma}}_{k=0} \rangle = \langle \sum_j \vec{\sigma}_j \rangle  = N \vec{S}$ is the normalized average total spin.

We now perform a time-dependent change of reference frame, introducing the right--handed Cartesian triple $(\hat{X},\hat{Y},\hat{Z})$ of unit vectors in the mobile frame ${\mathcal R}$, whose components in the original fixed frame $(\hat{x},\hat{y},\hat{z})$ are parameterized by the Euler angles $\theta=\theta(t)$ and $\phi=\phi(t)$ as
\begin{equation}
\label{eq:newbasis}
\hat{Z} \equiv 
\left( \begin{matrix}
\sin\theta \cos\phi \\
\sin\theta\sin\phi \\
\cos\theta
\end{matrix} \right) ; \qquad 
\hat{X} \equiv 
\left( \begin{matrix}
\cos\theta \cos\phi \\
\cos\theta\sin\phi \\
-\sin\theta
\end{matrix} \right) ; \qquad 
\hat{Y} \equiv 
\left( \begin{matrix}
-\sin\phi \\
\cos\phi \\
0
\end{matrix} \right) .
\end{equation}
%will be determined by requiring $\hat{z}$ to be the constantly aligned with the direction of the instantaneous total spin at all times.
Such a change of frame is implemented by the unitary operator $V=V(\theta,\phi)$, corresponding to a (time-dependent) global rotation defined by its action:
\begin{equation}
V \sigma_j^x V^{\dagger} = \hat{X} \cdot \vec{\sigma}_j \equiv \sigma_j^X, \qquad 
V \sigma_j^y V^{\dagger} = \hat{Y} \cdot \vec{\sigma}_j \equiv \sigma_j^Y, \qquad 
V \sigma_j^z V^{\dagger} = \hat{Z} \cdot \vec{\sigma}_j \equiv \sigma_j^Z,
\end{equation}
which can be written as $V = \exp\left(-i\phi\, s\sum_j\sigma_j^z\right) \, \exp\left(-i\theta\, s\sum_j\sigma_j^y\right)$.
The equations of motion of the components $\sigma_j^{\alpha}$ of the spins,
with $\alpha \in \{ X,Y,Z\}$, in the mobile frame $\mathcal R$ read
\begin{equation}
\frac{d}{dt} \sigma_j^{\alpha} =  \frac{1}{i}\big[\sigma_j^{\alpha},\widetilde{H}\big],
\end{equation}
with the modified Hamiltonian
\begin{equation}
\label{eq:inertial}
%H \quad \mapsto \quad 
\widetilde{H} \equiv  H + i V \dot{V}^{\dagger},
\end{equation}
which includes an additional $V$-dependent term, in analogy with the emergence 
%analogous to the appearance 
of apparent forces in classical mechanics when rotating coordinates are introduced.
In our case, a simple calculation %for this term 
shows that
\beq
i V \dot{V}^{\dagger} = - s \, \vec{\omega}(t) \cdot %\left( 
\sum_j \vec{\sigma}_j, % \right),
\eeq
where $\omega^X \equiv \hat{X}\cdot\vec{\omega} = -\sin\theta \; \dot{\phi}$, 
$\omega^Y \equiv \hat{Y}\cdot\vec{\omega} =  \dot{\theta}$,  and
$\omega^Z \equiv \hat{Z}\cdot\vec{\omega} = \cos\theta \; \dot{\phi}$.
%
%\beq
%\omega^x \equiv \hat{x}\cdot\vec{\omega} = -\sin\theta \; \dot{\phi}, \qquad 
%\omega^y \equiv \hat{y}\cdot\vec{\omega} =  \dot{\theta}, \qquad 
%\omega^z \equiv \hat{z}\cdot\vec{\omega} = \cos\theta \; \dot{\phi}.
%\eeq
Again, in analogy with classical mechanics, this latter result can be seen as a generalization of Larmor's theorem: the effect of the rotation of a reference frame is equivalent to the presence of a time-dependent external magnetic field.

The spins on the lattice can now be decomposed on the basis of $\mathcal R$ as
\begin{equation}
\vec{\sigma}_j =  \hat{X} \, \sigma_j^X \, + \hat{Y} \, \sigma_j^Y   + \hat{Z} \,  \sigma_j^Z .
\end{equation}
Accordingly, the Hamiltonian Eq.~\eqref{eq:Hs_fourierbar}, in the mobile frame $\mathcal R$, can be written as
\beq
\label{eq:Hrotatedframe}
\begin{split}
\frac{\widetilde{H}}{N} = &- g \left[ \left(\hat{X}\cdot\hat{z}\right) \frac{\tilde{\sigma}_{0}^X}{N} + \left(\hat{Y}\cdot\hat{z}\right) \frac{\tilde{\sigma}_{0}^Y}{N} +\left(\hat{Z}\cdot\hat{z}\right) \frac{\tilde{\sigma}_{0}^Z}{N}  \right]
  - \bar{\lambda}\left[ \left(\hat{X}\cdot\hat{x}\right) \frac{\tilde{\sigma}_{0}^X}{N} + \left(\hat{Y}\cdot\hat{x}\right) \frac{\tilde{\sigma}_{0}^Y}{N} +\left(\hat{Z}\cdot\hat{x}\right) \frac{\tilde{\sigma}_{0}^Z}{N}  \right]^2 \\
&- J \sum_{k\ne0} \cos k\left[ \left(\hat{X}\cdot\hat{x}\right) \frac{\tilde{\sigma}_{k}^X}{N} + \left(\hat{Y}\cdot\hat{x}\right) \frac{\tilde{\sigma}_{k}^Y}{N} +\left(\hat{Z}\cdot\hat{x}\right) \frac{\tilde{\sigma}_{k}^Z}{N}  \right]\left[ \left(\hat{X}\cdot\hat{x}\right) \frac{\tilde{\sigma}_{-k}^X}{N} + \left(\hat{Y}\cdot\hat{x}\right) \frac{\tilde{\sigma}_{-k}^Y}{N} +\left(\hat{Z}\cdot\hat{x}\right) \frac{\tilde{\sigma}_{-k}^Z}{N}  \right]   \\ 
&+\sin\theta \; s\dot{\phi} \; \frac{\tilde{\sigma}_0^X}{N} - s\dot{\theta} \; \frac{\tilde{\sigma}_0^Y}{N} - \cos\theta \; s\dot{\phi} \; \frac{\tilde{\sigma}_0^Z}{N} ,
\end{split}
\eeq
where the last line takes into account the additional term in Eq.~\eqref{eq:inertial}, while the components of the unit vectors in $\mathcal R$ with respect to those of the fixed frame are given in Eq.~\eqref{eq:newbasis} --- for instance, $\hat{X}\cdot\hat{z} = -\sin\theta$, $\hat{Y}\cdot\hat{z}=0$, $\hat{Z}\cdot\hat{z}=\cos\theta$, and analogously for all the other components.

\subsection{Time-dependent Holstein-Primakoff transformation}

In the rotating frame $\mathcal R$ with unit vectors $(\hat{X},\hat{Y},\hat{Z})$, we introduce the spin-wave canonical variables via the Holstein--Primakoff transformation
\begin{equation}
\label{eq:approxH-P}
\left\{
\begin{split}
\sigma_i^X &=\frac { q_i}{\sqrt{s}}  + \mathcal{O}\left((q_i,p_i/\sqrt{s})^3\right), \\
\sigma_i^Y  &= \frac { p_i}{\sqrt{s}}+ \mathcal{O}\left((q_i,p_i/\sqrt{s})^3\right), \\
%\mathcal{O}\bigg(\frac{q_j,p_j}{\sqrt{s}}\bigg)^3  \\
\sigma_i^Z  &= 1-\frac{b^{\dagger}_i b_i}{s} \equiv 1 - \frac{q_i^2+p_i^2-1}{2s},
\end{split} 
\right.
\end{equation}
where $q_i$ and  $p_i$ are the conjugate canonical variables representing small deviations of the spin away from the $\hat{Z}$-axis, and along the  directions $\hat{X}$ and $\hat{Y}$, respectively, while  $b_j =  (q_j + i p_j)/\sqrt{2}$ in our notation. The formal expansion of the operators $\sigma_i^{X,Y,Z}$ is in powers of $q_j/\sqrt{s}$ and $p_j/\sqrt{s}$ and in Eqs.~\eqref{eq:approxH-P} we have retained the leading orders.
Accordingly,  defining the Fourier space coordinates
 $\tilde{q}_k= N^{-1/2} \sum_j e^{-ikj}q_j$ and $\tilde{p}_k= N^{-1/2} \sum_j e^{-ikj}p_j$, we get
\begin{equation}
\label{eq:approxH-Pfourier}
\left\{
\begin{split}
\frac{\tilde{\sigma}_{k}^X}{N} =&  \frac{\tilde{q}_{k}}{\sqrt{Ns}} +  \mathcal{O}\left( (\tilde{q},\tilde{p}/\sqrt{Ns})^3 \right),  \\
\frac{\tilde{\sigma}_{k}^Y}{N} =&  \frac{\tilde{p}_{k}}{\sqrt{Ns}} + \mathcal{O}\left( (\tilde{q},\tilde{p}/\sqrt{Ns})^3  \right), \\
\frac{\tilde{\sigma}_{k}^Z}{N}   =& \delta_{k,0}-\sum_{k'}\frac{\tilde{q}_{k'}\tilde{q}_{k-k'}+\tilde{p}_{k'}\tilde{p}_{k-k'}-\delta_{k,0}}{2(\sqrt{Ns})^2}  .
 \end{split} 
 \right.
 \end{equation}
%
%
%
%
%Notice that in the classical limit, formally $s\to\infty$, the zero-point contributions vanish and equations \eqref{eq:approxH-P}, (and hence \eqref{eq:approxH-PfourierX}, \eqref{eq:approxH-PfourierY}, \eqref{eq:approxH-PfourierZ}) become exactly expansions in powers of $q' \equiv q/\sqrt{s}$ and $ p' \equiv p/\sqrt{s} $.
%
%
%
At the lowest non-trivial order in the density of spin-waves --- controlled by $\epsilon(t)$ introduced below, see Eq.~\eqref{eq:epsilondef} ---
a straightforward calculation shows that the modulus $|\vec{\Sigma}_\text{tot}|$ of the total spin 
\beq
\vec{\Sigma}_{\text{tot}} \equiv  s \sum_j \vec{\sigma}_j,
\eeq
is given by
\begin{equation}
\label{eq:absStot}
\big\lvert \vec{\Sigma}_{\text{tot}} \big\rvert^2= \left( Ns-\sum_{k\ne0} \tilde{b}_k^{\dagger} \tilde{b}_k \right) \left( Ns-\sum_{k\ne0} \tilde{b}_k^{\dagger} \tilde{b}_k  +1\right),
%\hat{z}\cdot \vec{S}_{\text{tot}}= \frac{1}{i} \Big[\hat{x}\cdot \vec{S}_{\text{tot}}, \hat{y}\cdot \vec{S}_{\text{tot}}\Big] = Ns - \sum_{k} \tilde{b}_k^{\dagger} \tilde{b}_k 
%\label{eq:Stotz}
\end{equation}
where $\tilde{b}_{k}\equiv  N^{-1/2} \sum_j e^{-ikj} b_j $.
Note that all the excitations with momenta $k$ decrease the total spin projection along the instantaneous direction $\hat{Z}$ of the ``vacuum" $\langle \vec{\Sigma}\rangle$ (from Eq.~\eqref{eq:approxH-Pfourier}, one gets $\Sigma_{\text{tot}}^Z \equiv s \tilde{\sigma}_0^Z = Ns - \sum_k \tilde{b}_k^{\dagger} \tilde{b}_k$), 
but only excitations with $k\ne0$ decrease the modulus $\big\lvert \vec{\Sigma}_{\text{tot}} \big\rvert$ of the total spin, see Eq.~\eqref{eq:absStot}.
In other words, the spin-wave operators with $k=0$ dictate the motion of the spin-wave vacuum.

In order to derive the equations of motion for the spins, one should substitute the expansions \eqref{eq:approxH-Pfourier} into the Hamiltonian \eqref{eq:Hrotatedframe} and calculate its commutator with the canonical variables $\tilde{q}_{k}$, $\tilde{p}_{k}$. 
Truncating these expansions at the lowest orders is justified as long as:
\begin{enumerate}
\item The time-dependent angles $\theta(t)$ and $\phi(t)$ which control the rotating frame $\mathcal R$ are chosen such that the total spin $\vec \Sigma$ remains constantly aligned with the rotating $\hat{Z}$ axis, i.e., $\forall t>0$,
\begin{equation}
\label{eq:timedepvacuum}
\big\langle \tilde{\sigma}_{0}^X(t) \big\rangle = \big\langle  \tilde{\sigma}_{0}^Y(t) \big\rangle = 0 \quad \text{or, equivalently,} \quad S_X(t)= S_Y(t)= 0.
\end{equation}
\item The spin-waves population remains small, i.e.,
\begin{equation}\label{eq:smallspin}
\sum_{k\ne0}\langle n_k(t) \rangle = \sum_{k\ne0} \Big\langle \frac{ \tilde{q}_{k}(t)\tilde{q}_{-k}(t)+\tilde{p}_{k}(t)\tilde{p}_{-k}(t)-1 }{2} \Big\rangle  \ll Ns,
\end{equation}
where we defined
\beq
\label{eq:def-nk}
 n_k \equiv \tilde{b}_k^\dagger \tilde{b}_k.
\eeq
\end{enumerate}
The first condition is fulfilled by requiring the equations of motion  for $S_X(t)$ and $S_Y(t)$ to be trivially
\beq
\label{eq:thetaphi}
\frac{d}{dt}\big\langle \tilde{\sigma}_{0}^X(t) \big\rangle = \frac{d}{dt} \big\langle  \tilde{\sigma}_{0}^Y(t) \big\rangle = 0, \quad\mbox{with}\quad \big\langle \tilde{\sigma}_{0}^X(t=0) \big\rangle = \big\langle  \tilde{\sigma}_{0}^Y(t=0) \big\rangle = 0.
\eeq
These two equations  determine the motion of the rotating frame, i.e., the time-evolution of the Euler angles $\theta(t)$ and $\phi(t)$. The validity of the second 
condition, i.e., of Eq.~\eqref{eq:smallspin}, can be checked 
by monitoring the time evolution of the total spin-wave density
\begin{equation}
\label{eq:epsilondef}
\epsilon(t) \equiv  \frac{1}{Ns} \sum_{k\ne0}\big\langle n_k (t) \big\rangle = \frac{1}{Ns} \sum_{k\ne0}\bigg\langle\frac{\tilde{q}_{k}(t)\tilde{q}_{-k}(t)+\tilde{p}_{k}(t)\tilde{p}_{-k}(t)-1}{2}\bigg\rangle,
\end{equation}
which also quantifies to the deviation of the total spin $\big\lvert \vec{\Sigma}_{\text{tot}} \big\rvert$ from its maximal value $Ns$ according to Eq.~\eqref{eq:absStot}. 
The approximations introduced above are consistent as long as $\epsilon(t)\ll1$; if the  total spin-wave density happens to become larger, then higher-order terms in the canonical spin-wave coordinates are expected to contribute to the dynamics at longer times.

\subsection{Equations of motion within the Gaussian approximation}

The simplest non-trivial approximation consists in treating quantum fluctuations as being harmonic, i.e., within the Gaussian approximation; this means that the expansion in Eq.~\eqref{eq:approxH-Pfourier} is substituted into the Hamiltonian \eqref{eq:Hrotatedframe} and only the \emph{linear} terms in the vacuum coordinates $(\tilde{q}_0,\tilde{p}_0$) and the \emph{quadratic} terms in the spatial fluctuation coordinates $(\tilde{q}_k,\tilde{p}_k)$ with $k\ne0$ are kept. 

Let us first discuss  the  mean-field case $J=0$ (discussed in Ref.~[21] of the Letter) in which the Hamiltonian $H_{J=0}$ is a function of the total spin $\tilde{\vec{\sigma}}_0$ only. Accordingly, the  modes $\tilde{q}_k$ and $\tilde{p}_k$ with $k\ne0$ enter the Hamiltonian within the Gaussian approximation only via
\begin{equation}
\frac{\tilde{\sigma}^Z_0}{N} =   1- \frac{1}{Ns} \sum_{k\ne0}  \frac{\tilde{q}_{k}(t)\tilde{q}_{-k}(t)+\tilde{p}_{k}(t)\tilde{p}_{-k}(t)-1}{2},
\end{equation}
i.e., $\langle \tilde{\sigma}^Z_0  \rangle = N(1-\epsilon)$, while each spin-wave number $n_k$ (see Eq.~\eqref{eq:def-nk}) is a constant of motion.
In fact, $[ n_k \, , \, \tilde{\vec{\sigma}}_{0}]=0$ implies  $[ n_k  \, , \, H_{J=0}]=0$ and therefore, from Eq.~\eqref{eq:epsilondef}, 
\beq
\quad \frac{d}{dt}\epsilon = 0, 
\eeq
which corresponds to the conservation of the total spin (see Eq.~\eqref{eq:absStot}).
Imposing the condition \eqref{eq:timedepvacuum}, we get the classical evolution equation of the total spin in the present mean-field case $J=0$, which coincides with the equations of motion found in Ref.~[21] of the Letter.

Consider, now, the case $J\neq 0$ which introduces the short-range interaction term $U$ 
(corresponding to the second line of Eq.~\eqref{eq:Hrotatedframe})
in the infinite-range Hamiltonian discussed above;
in particular,  our main goal consists in understanding 
the influence of $U$ on the mean-field dynamics discussed above.%
In this respect it is convenient to write $U$ as $U=U_1+U_2+U_3$ where
\begin{eqnarray}
U_1&=&- J \sum_{k\ne0} \cos k \left[ \left(\hat{X}\cdot\hat{x}\right)^2 \frac{\tilde{\sigma}_{k}^X}{N}\frac{ \tilde{\sigma}_{-k}^X}{N}
+\left(\hat{Y}\cdot\hat{x}\right)^2 \frac{\tilde{\sigma}_{k}^Y}{N}\frac{ \tilde{\sigma}_{-k}^Y}{N}
+ \left(\hat{X}\cdot\hat{x}\right) \left(\hat{Y}\cdot\hat{x}\right) \left( \frac{\tilde{\sigma}_{k}^X}{N}\frac{ \tilde{\sigma}_{-k}^Y}{N} +  \frac{\tilde{\sigma}_{k}^Y}{N}\frac{ \tilde{\sigma}_{-k}^X}{N} \right) \right],\\
U_2&=& -J \sum_{k\ne0} \cos k \left[
\left(\hat{X}\cdot\hat{x}\right)\left(\hat{Z}\cdot\hat{x}\right) \left( \frac{\tilde{\sigma}_{k}^X}{N}\frac{ \tilde{\sigma}_{-k}^Z}{N} + \frac{\tilde{\sigma}_{k}^Z}{N}\frac{ \tilde{\sigma}_{-k}^X}{N} \right)
+\left(\hat{Y}\cdot\hat{x}\right)\left(\hat{Z}\cdot\hat{x}\right) \left( \frac{\tilde{\sigma}_{k}^Y}{N}\frac{ \tilde{\sigma}_{-k}^Z}{N} +  \frac{\tilde{\sigma}_{k}^Z}{N}\frac{ \tilde{\sigma}_{-k}^Y}{N} \right) \right],\\
U_3&=&-J \sum_{k\ne0} \cos k  \left(\hat{Z}\cdot\hat{x}\right)^2 \frac{\tilde{\sigma}_{k}^Z}{N}\frac{ \tilde{\sigma}_{-k}^Z}{N} .
\end{eqnarray}
Expanding $U_{1,2,3}$  by means of Eqs.~\eqref{eq:approxH-Pfourier}, it is easy to realise  that 
$U_1$  gives rise to quadratic terms in $(\tilde{q}_k,\tilde{p}_k)$; $U_3$ gives rise to a quartic (i.e., two-body)
interaction among spin-waves, which is therefore negligible in the low-density limit $\epsilon\ll1$; $U_2$, instead,  generates contributions which are simultaneously linear in the vacuum coordinates $(\tilde{q}_0, \tilde{p}_0)$ and quadratic in the spin-waves modes $(\tilde{q}_k, \tilde{p}_k)$. These terms therefore couple the motion of the vacuum with the spin-waves motion at the lowest non-trivial order: accounting for them is crucial in order to understand the modifications (which we refer to as ``feedback'') to the mean-field motion caused by the quantum fluctuations, which is the goal of our work.

The equations of motion of the mobile frame (i.e., by construction, of the collective spin), 
including the feedback of quantum fluctuations due to $U$,  are found by imposing Eq. \eqref{eq:thetaphi}, and read
\begin{equation}
\label{eq:vacuummotion}
\left\{
\begin{split}
 \frac{d}{dt}\theta =& + 4 \bar{\lambda} (1-\epsilon)  \sin\theta \cos\phi \sin\phi \\
                         & - 4J  \bigg( \frac{1}{Ns} \sum_{k\ne0} \cos k \;  \left\langle \tilde{p}_k \tilde{p}_{-k} \right\rangle \bigg)  \sin\theta \cos\phi \sin\phi \\
                         & +4J \Bigg( \frac{1}{Ns} \sum_{k\ne0} \cos k   \frac{\left \langle \tilde{q}_k \tilde{p}_{-k} + \tilde{p}_k \tilde{q}_{-k}\right\rangle}{2} \Bigg) \cos\theta  \sin\theta \cos^2\phi, \\
 \frac{d}{dt}\phi =& -2g + 4 \bar{\lambda} (1-\epsilon)   \cos\theta \cos^2\phi  \\
                                          & -4J \bigg( \frac{1}{Ns} \sum_{k\ne0} \cos k \; \left\langle  \tilde{q}_k \tilde{q}_{-k} \right\rangle \bigg) \cos\theta \cos^2\phi  \\
                                          & +4J \Bigg( \frac{1}{Ns} \sum_{k\ne0} \cos k   \frac{\left\langle\tilde{q}_k \tilde{p}_{-k} + \tilde{p}_k \tilde{q}_{-k}\right\rangle}{2} \Bigg) \sin\phi \cos\phi  , 
\end{split}
\right.
\end{equation}
where $\epsilon$ is defined in Eq.~\eqref{eq:epsilondef}. The equal-time correlation functions appearing in these equations are
\begin{equation}
\label{eq:deltadef}
\begin{split}
\Delta^{qq}_k (t) &\equiv  \left\langle \tilde{q}_k(t) \tilde{q}_{-k}(t) \right\rangle ,\\
\Delta^{pp}_k (t) &\equiv  \left\langle \tilde{p}_k(t) \tilde{p}_{-k}(t) \right\rangle, \\
\Delta^{qp}_k (t) &\equiv \frac{1}{2}\left\langle \tilde{q}_k(t) \tilde{p}_{-k}(t) + \tilde{p}_k(t) \tilde{q}_{-k}(t) \right\rangle.
\end{split}
\end{equation}
Using now the equations of motion for the spin-waves coordinates, found by  computing their commutators with the Hamiltonian $\tilde{H}$,
\begin{equation}
\label{eq:swmotion}
\left\{
\begin{split}
 \frac{d}{dt} \tilde{q}_k 
                                   =& +4\bar{\lambda}  \cos^2 \phi \; \tilde{p}_k 
                                  % \\ & 
                                   -4J \cos k \, \sin^2\phi \; \tilde{p}_k +4J\cos k \, \cos\theta\cos\phi\sin\phi  \; \tilde{q}_k , \\  
 \frac{d}{dt} \tilde{p}_k
                                   =& -4\bar{\lambda}  \cos^2 \phi \; \tilde{q}_k %\\& 
                                   +4J \cos k \, \cos^2\theta\cos^2\phi \; \tilde{q}_k -4J\cos k \, \cos\theta\cos\phi\sin\phi  \; \tilde{p}_k ,
\end{split}
\right.
\end{equation}
one obtains
\begin{equation}
\label{eq:swmotiondelta}
\left\{
\begin{split}
\frac{d}{dt}\Delta^{qq}_k  = \, & 8J\cos k\,  \cos\theta \cos\phi\sin\phi \,  \Delta^{qq}_k %\\& 
+8\left(\bar{\lambda} \cos^2\phi -J \cos k\,  \sin^2\phi \right)\, \Delta^{qp}_k ,\\
\frac{d}{dt}\Delta^{qp}_k  = & -4\left(\bar{\lambda} \cos^2\phi - J    \cos k \,   \cos^2\theta \cos^2\phi \right) \Delta^{qq}_k %\\& 
+ 4 \left(\bar{\lambda} \cos^2\phi -J\cos k \, \sin^2\phi \right) \Delta^{pp}_k ,\\
\frac{d}{dt}\Delta^{pp}_k  = & -8\left( \bar{\lambda} \cos^2\phi - J\cos k \, \cos^2\theta \cos^2\phi \right) \Delta^{qp}_k %\\& 
-8 J\cos k \,  \cos\theta \cos\phi\sin\phi \,  \Delta^{pp}_k.
\end{split}
\right.
\end{equation}
Note that the spin-waves --- and therefore their correlators $\Delta_k^{\alpha\beta}$ with $\alpha$, $\beta \in \{p,q\}$ --- have a dynamics even for $J=0$ and $\lambda\ne0$, though trivial, as it amounts at  conserving the spin-wave occupation numbers $n_k$ in Eq.~\eqref{eq:def-nk}. These, numbers, instead,  are not conserved for $J \neq 0$. 
In addition, the equations of motion written above are not actually independent because a Gaussian wavefunction such as the one of the spin-waves within the present harmonic approximation is completely specified by \emph{two} parameters rather than the \emph{three} $\Delta^{qq}$, $\Delta^{qp}$, and $\Delta^{pp}$. In fact, the latter quantities are actually related by the condition  
\beq
4 \left(\Delta^{qp}_k\right)^2 = 4 \Delta^{qq}_k \Delta^{pp}_k - 1,
\eeq 
which is satisfied at all times and for all values of $k$.

The ``feedback'' terms $\propto J$ appearing in the equations of motion \eqref{eq:vacuummotion} of the vacuum are of the form
\beq
\label{eq-sm:def-delta}
\delta^{\alpha\beta} \equiv \frac{1}{Ns} \sum_{k\ne0}  \Delta^{\alpha\beta}_k \cos k,
\eeq
hence Eq.s \eqref{eq:vacuummotion} can be written as
\begin{equation}
\label{eq:vacuummotiondelta}
\left\{
\begin{split}
&\dot{\theta} = 4 \bar{\lambda} (1-\epsilon)  \sin\theta \cos\phi \sin\phi  - 4J   \delta^{pp}  \sin\theta \cos\phi \sin\phi 
                         +4J  \delta^{qp} \cos\theta  \sin\theta \cos^2\phi, \\
&\dot{\phi} = -2g + 4 \bar{\lambda} (1-\epsilon)   \cos\theta \cos^2\phi  
                                    -4J\delta^{qq} \cos\theta \cos^2\phi 
                                          +4J \delta^{qp} \sin\phi \cos\phi ,
\end{split}
\right.
\end{equation}
where, from Eqs. \eqref{eq:epsilondef} and \eqref{eq:deltadef}, 
\begin{equation}
\epsilon \equiv \frac{1}{Ns} \sum_{k\ne0} \frac{\Delta^{qq}_k + \Delta^{pp}_k -1}{2} 
\end{equation}
(cf. Eq. (3) in the Letter).
Equations \eqref{eq:vacuummotiondelta} and \eqref{eq:swmotiondelta} provide  the final system of $2N$ coupled ordinary differential equations which yield the post-quench dynamics at linear order in the spin-wave density $\epsilon$, where we recall that $N$ is the number of spins on the lattice, i.e., the number of possible discrete momenta $k$. 
These equations are expected not to provide accurate results whenever $\epsilon(t)$ increases and approaches values of  order $1$. 
Note that, as $J\to0$, the motion of $\theta$ and $\phi$ decouples from the quantum fluctuations and we retrieve the mean-field limit; the same happens of course in the formal classical limit $s\to\infty$.

In order to solve simultaneously the evolution equations~\eqref{eq:swmotiondelta} and \eqref{eq:vacuummotiondelta}, we need to prescribe the initial conditions, which depend on the specific quench under consideration.
For quenches of the Hamiltonian $H$ in Eq.~\eqref{eq:Hs_fourierbar} originating from the ground state corresponding to $g=g_0$ and $J=0$, the initial state is the perfectly coherent state with all the spins pointing in the direction given by the minimum of the classical pre-quench Hamiltonian. 
Accordingly, the initial conditions turn out to be
\begin{equation}\label{eq:initialcond}
%\begin{split}
%\theta(t=0) = & \arccos(g_0/(2\lambda)) ; \\
%\phi(t=0) = & 0 ; \\
%\Delta^{qq}_k(t=0) = & 1/2 ; \\
%\Delta^{pp}_k(t=0) = & \frac{1}{2} ; \\
%\Delta^{qp}_k(t=0) = & 0;
%\end{split}
\theta(t=0) = \arccos(g_0/(2\lambda)), \quad \phi(t=0) =  0, \quad \Delta^{qq}_k(t=0) = \Delta^{pp}_k(t=0) =1/2,\quad\mbox{and}\quad \Delta^{qp}_k(t=0) = 0,
\end{equation}
for all $k\ne0$; in particular, $\epsilon(t=0)=0$. In the Letter we always consider quenches with $g_0=0$, for which the  pre-quench  value of $J$ is actually inconsequential and it may be taken equal to the post-quench value,  so that the only parameter affected by the quench is $g$.
As a final remark, we note that our approach can be easily generalized in order to deal with a more general class of spin  models on an arbitrary graph structure (including higher spatial dimensions) and with arbitrary spin-spin couplings.

\section{\large 2. Perturbation theory at the dynamical critical point: evaluation of $\delta^{qq}(t)$}

\label{jamir}

In this section we discuss the form of the quantum feedback terms $\delta^{\alpha\beta}$ (see Eqs.~\eqref{eq-sm:def-delta} and \eqref{eq:deltadef}) and their long-time  behavior in a regime which is analytically tractable.
In particular, we consider the unperturbed dynamics with $J=0$  and at the dynamical critical point, as a reference for introducing a  leading-order  perturbation theory  in $J\ll g$, $\lambda$. 

The corresponding mean-field dynamics of the angles $\theta(t)$ and $\phi(t)$ reads $\tan\phi(t)=-\tanh(\bar{\lambda}t)$ and $\cos\theta(t) =\tanh^2(\bar{\lambda}t)$ (see, for instance, Ref. [21] in the Letter).
Inserting these expressions into the system~\eqref{eq:swmotiondelta} and considering the leading-order contributions in an expansion at long times with $\bar{\lambda}t\gg1$, the evolution equations take the simpler form
\begin{equation}\label{eq:systempert}
\left\{
\begin{split}
&\frac{d}{dt}(\Delta^{qq}_k+\Delta^{pp}_k)  =   -2J\cos k\,(\Delta^{qq}_k-\Delta^{pp}_k), \\
&\frac{d}{dt}(\Delta^{qq}_k-\Delta^{pp}_k)  =   -2J\cos k\,(\Delta^{qq}_k+\Delta^{pp}_k) +4(\bar{\lambda}-J\cos k)\Delta^{pq}_k,\\
&\frac{d}{dt}\Delta^{pq}_k  =  -(\bar{\lambda}-J\cos k)(\Delta^{qq}_k-\Delta^{pp}_k),
\end{split}
\right.
\end{equation}
where, for convenience, we have chosen $\Delta^{qq}_k \pm \Delta^{pp}_k$ instead of $\Delta^{qq}_k$ and $\Delta^{pp}_k$, as new active variables.
Assuming generic initial conditions of the form $(\Delta^{qq}_k+\Delta^{pp}_k)|_{t=0}=a_1$, $(\Delta^{qq}_k-\Delta^{pp}_k)|_{t=0}=a_2$, $(\Delta^{pq}_k)|_{t=0}=a_3$, we solve  the dynamics prescribed by Eq.~\eqref{eq:systempert} at the lowest  non-trivial order in $J$. 
Such a solution allows us to calculate explicitly, for instance, the quantum feedback $\delta^{qq}(t)$ (see Eqs.~\eqref{eq-sm:def-delta} and \eqref{eq:deltadef})  in the thermodynamic limit ($\sum_{k\neq0}...\to N\int^{2\pi}_0\frac{dk}{2\pi}...$)
\begin{equation}
\begin{split}
\delta^{qq}(t)=&\frac{2J}{N}\sum_{k\neq0}\cos k\,\Delta^{qq}_k=\frac{J}{\pi }\int^{2\pi}_0dk\cos k\,\Delta^{qq}_k=\\
&=\frac{\pi}{4\bar{\lambda}t}\left\{- (a_1+a_2) \left[ 2Jt I_0(2Jt)-I_1(2Jt)\right] \sin(2\bar{\lambda}t)
- a_2 \left[-2\bar{\lambda}t I_1(2Jt)\right]\sin(2\bar{\lambda}t)\right.\\
&\left.+2 a_3\left(-Jt+ \left[2JtI_0(2Jt) - I_1(2Jt)\right]\left[\cos(2\bar{\lambda}t)-\sin(2\bar{\lambda}t)\right]
- 2\bar\lambda t I_1(2Jt)\cos(2\bar{\lambda}t) \right)\right\}, 
%&=\frac{\pi}{4gt}(-(2Jt I_0(2Jt)-I_1(2Jt))\sin(2gt)a_1-(2Jt I_0(2Jt)-(1+2gt)I_1(2Jt))\sin(2gt)a_2+\\
%&+2(-Jt+2JtI_0(2Jt)(\cos(2gt)-\sin(2gt))+I_1(2Jt)(-(1+2gt)\cos(2gt)+\sin(2gt)))a_3), 
\end{split}
\end{equation}
where $I_n(x)$ is the Bessel function of first kind, with index $n$ and argument $x$.
In the long-time limit $Jt\gg1$, one can employ the asymptotic expansion of $I_n(x)$ for  $x\gg1$, specifically for $n=0,1$,
\begin{equation}
I_{(0,1)}(x)\sim \pm\cos\left(\frac{\pi}{4}\mp x\right)\sqrt{\frac{2}{\pi x}}+\mathcal{O}\left(\frac{1}{x}\right) ,
\end{equation}
in order to find that, in addition to a constant term, $\delta^{qq}(t)$ decays as $(Jt)^{-1/2}$ modulated by oscillatory terms of the form $\cos(\pi/4\mp2(\bar{\lambda}\pm J)t)$ and $\sin(\pi/4\pm2(\bar{\lambda}\mp J)t)$, which result from  beats of the two frequencies $2J$ and $2\bar{\lambda}$. Analogous qualitative results hold for $\delta^{pp}(t)$ and $\delta^{qp}(t)$.

\section{\large 3. Convergence of the MPS-TDVP}

As discussed in the Letter, we investigated the dynamical behavior of the system under study also for values of the parameters at which the spin-wave approximation discussed in the previous sections is not expected to be accurate.
In this case, we used the matrix product state
time-dependent variational principle (MPS-TDVP~[23,~24])
and we assessed its viability for investigating the dynamics of the average longitudinal magnetization $S_x(t)$ at long times by studying  its finite-size scaling, i.e., how it changes upon increasing the systems size $N$. 
In addition, for each value of $N$ investigated here,  we also studied the dependence of the numerical result on the bond dimension $D$. 
For all simulations, we used a fourth-order integrator with time step 0.02
and the MPS-TDVP formulation developed in ~[23,~24].

In Fig.~\ref{fig:ndconv} we compare the time evolution of the order parameter $S_x(t)$,
obtained by using two different bond dimensions $D=80$ and $D=128$, three
values of the system size $N=100,~200,~400$, and three post-quench values of $g=0.5$, 0.83 (corresponding to the ferromagnetic phase), and 1.33 (paramagnetic phase), while keeping fixed $J=2/3$ and $\lambda=1/3$. 
The systematic error of this approach can be estimated as
the difference between curves which differ only for the value of $D$ and it turns out to be of the order of 
10\% at intermediate times; however, we consistently observe faster convergence in the time averaged order
parameter $\bar S_x$, for which the estimated error is few percents.
Interestingly, we observe faster convergence with the bond dimension for larger systems. 
%
%
%%%%%%%%%%%%%%%%%%%%%
%   FIG 1  %%%%%%%%%%%%%%
%%%%%%%%%%%%%%%%%%%%%
\begin{figure}[t!]
  \includegraphics[width=18cm]{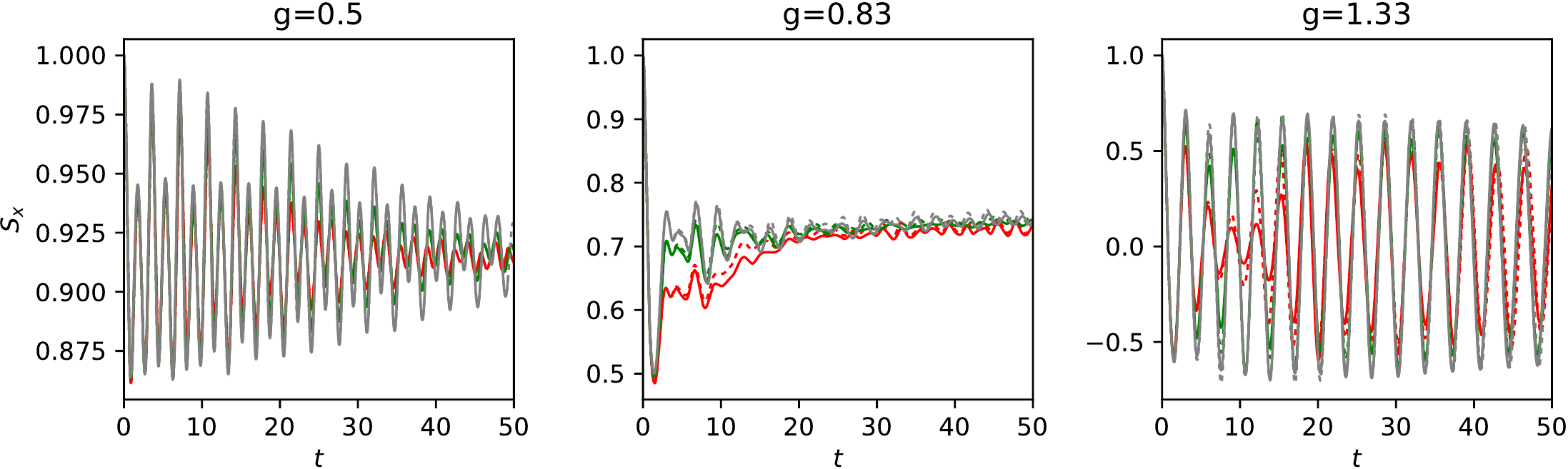}
  \caption{Dependence of the evolution of $S_x(t)$ obtained from  MPS-TDVP
    simulations on the system size $N=100$~(red),~200~(green)~400~(gray) and the
    bond dimension $D=80$~(dashed line)~128~(solid line), for three values of the post-quench parameter $g=0.5$, 0.83, 1.33, and fixed $J=2/3$ and $\lambda=1/3$. 
  In all cases we observe that upon varying the bond dimension $D$ the resulting
  evolution is essentially unaffected. In addition, we observe that in the
  ferromagnetic region (g=0.5,~0.83) the order
  parameter increases with increasing the system size indicating a non-vanishing
  value in the thermodynamic limit.
}
  \label{fig:ndconv}
\end{figure}
%%%%%%%%%%%%%%%%%%%%%
%%%%%%%%%%%%%%%%%%%%%

The behavior of the dynamics of the order parameter $S_x$ in the limit $N\to\infty$ is inferred here 
from observing how it changes upon increasing $N$, see Fig.~\ref{fig:ndconv}.
While at intermediate times the finite-size effects are still large, the long-time averaged observables such as 
$\bar S_x$ converge much faster to the thermodynamic limit, as no changes are observed upon further increasing $N$.
Moreover, in the ferromagnetic regime we observe that the order parameter increases upon increasing 
the system size $N$, indicating a non-vanishing order parameter as $N\rightarrow\infty$. This last observation is valid also in the case where the final magnetization is
reversed with respect to the initial state.

Similar fast convergence of $\bar S_x$ is observed upon increasing the bond dimension $D$ and the system size $N$, also in the chaotic region with
initial and final magnetizations of opposite sign, as shown by the curves in
Fig.~\ref{fig:ndconv2}. In fact, while they display a significant dependence on
$N$ as a consequence of the chaotic behavior consistently observed in that
region of the parameter space, the time averaged order parameter $\bar S_x$ is eventually independent of the bond dimension $D$.

%%%%%%%%%%%%%%%%%%%%%
%   FIG 2  %%%%%%%%%%%%%%
%%%%%%%%%%%%%%%%%%%%%
\begin{figure}[t!]
  \includegraphics[width=9cm]{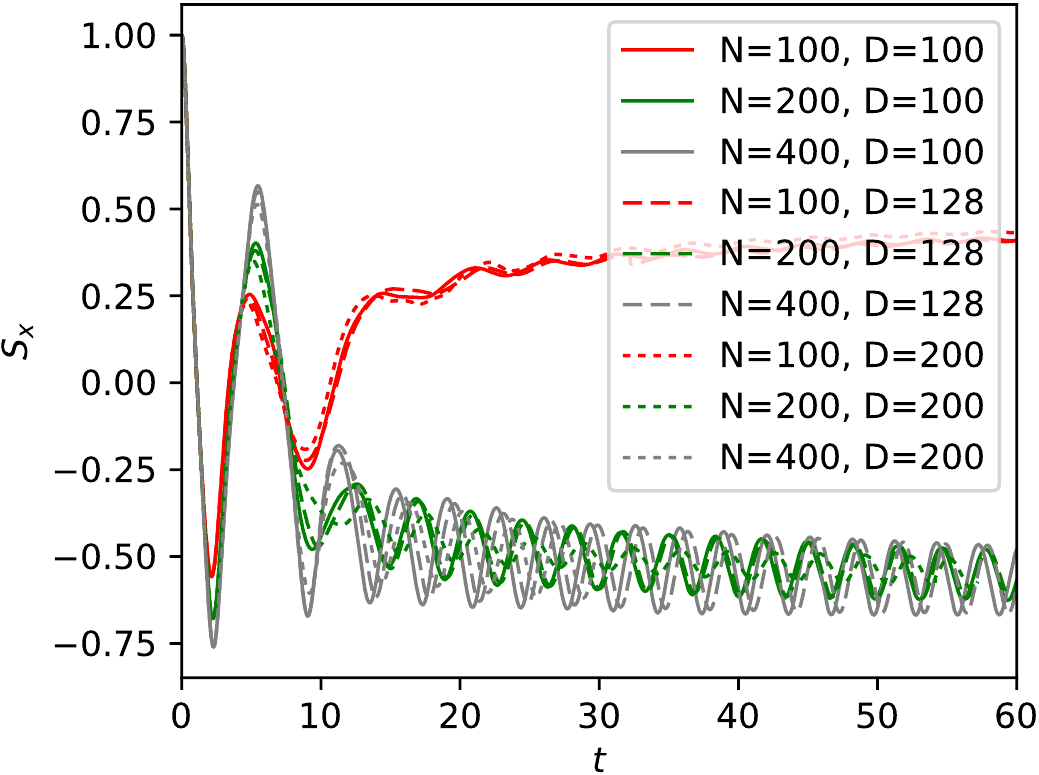}
  \caption{Dependence of the evolution of $S_x(t)$ obtained from MPS-TDVP simulations on the system size $N$ and the bond dimension $D$, for a value $g=1.1$ 
  of the field which turns out to correspond to the chaotic dynamical ferromagnetic region and fixed $J=1/2$ and $\lambda=1/2$. 
  While the system size $N$ has a significant effect on the evolution of $S_x$ and therefore on the value of $\bar S_x$ --- a fact which characterizes the chaotic dynamical ferromagnetic phase --- upon increasing the bond dimension $D$, no qualitative changes occur. 
    }
  \label{fig:ndconv2}
\end{figure}
%%%%%%%%%%%%%%%%%%%%%
%%%%%%%%%%%%%%%%%%%%%

%\bibliography{biblio}
%\begin{thebibliography}{20}
%\bibitem{kopietz} A. R\"uckriegel, A. Kreisel, and P. Kopietz, Phys. Rev. B
%85, 054422 (2012).
%\bibitem{tdvp1} C. Lubich, I. V. Oseledets, and B. Vandereycken, SIAM J. Numer. Anal 53, 917 (2015).
%\bibitem{tdvp2} J. Haegeman, C. Lubich, I. Oseledets, B. Vandereycken, and F. Verstraete, Phys. Rev. B 94, 165116 (2016).
% \bibitem{Sciolla2013} B. Sciolla and G. Biroli,  Phys. Rev. B {\bf 88}, 201110 (2013).
%\end{thebibliography}

\end{widetext}

\end{document}